{}
{}

\documentclass[11pt,a4paper]{article}

\newif\ifpublic\publictrue
\newif\iffancy\fancytrue

\usepackage[a4paper,text={160mm,247mm},centering]{geometry}
\usepackage{setspace}
\usepackage{amsmath,amssymb, amsfonts, geometry}
\makeatletter
\providecommand*{\shuffle}{%
  \mathbin{\mathpalette\shuffle@{}}%
}
\newcommand*{\shuffle@}[2]{%
  \sbox0{$#1\vcenter{}$}%
  \kern .15\ht0 
  \rlap{\vrule height .25\ht0 depth 0pt width 2.5\ht0}%
  \raise.1\ht0\hbox to 2.5\ht0{%
    \vrule height 1.75\ht0 depth -.1\ht0 width .17\ht0 %
    \hfill
    \vrule height 1.75\ht0 depth -.1\ht0 width .17\ht0 %
    \hfill
    \vrule height 1.75\ht0 depth -.1\ht0 width .17\ht0 %
  }%
  \kern .15\ht0 
}
\makeatother

\usepackage[pdfencoding=auto,bookmarks=true,hyperfigures=true]{hyperref}
\usepackage{graphicx}
\usepackage{float}
\usepackage[nosort]{cite}
\usepackage{lmodern}
\usepackage{amsbsy}
\usepackage[utf8]{inputenc}
\usepackage[font=small,labelfont=bf]{caption}
\usepackage{diagbox}
\usepackage{array}

\usepackage[usenames,dvipsnames]{xcolor}
\definecolor{dgreen}{rgb}{0,0.70,0.30}
\definecolor{gold}{rgb}{0.85,.66,0}
\definecolor{purple}{rgb}{1.0,0.3,0.6}

\usepackage{tikz}
\usetikzlibrary{calc} 
\usetikzlibrary{patterns} 
\usetikzlibrary{decorations.pathreplacing} 
\usetikzlibrary{decorations.markings} 
\usetikzlibrary{decorations.pathmorphing} 
\usetikzlibrary{positioning}

\iffancy

\def\showkeysrefformat#1{{\normalfont\tiny\ttfamily#1}}
\makeatletter
\def\SK@@ref#1>#2\SK@{%
 {\@inlabelfalse\leavevmode\vbox to\z@{%
 \vss\SK@refcolor\rlap{\vrule\raise .75em%
  \hbox{\showkeysrefformat{#2}}}}}}
\makeatother
\fi

\numberwithin{equation}{section}

\providecommand{\href}[2]{#2}

\makeatletter
\def\mr@ignsp#1 {\ifx\:#1\@empty\else #1\expandafter\mr@ignsp\fi}%
\newcommand{\multiref}[1]{\begingroup
\xdef\mr@no@sparg{\expandafter\mr@ignsp#1 \: }%
\def\mr@comma{}%
\@for\mr@refs:=\mr@no@sparg\do{\mr@comma\def\mr@comma{,}\ref{\mr@refs}}%
\endgroup}
\renewcommand{\eqref}[1]{(\multiref{#1})}
\makeatother

\makeatletter
\newcommand{\namedref}[2]{\hyperref[#2]{#1~\ref*{#2}}}
\newcommand{\secref}{\@ifstar{\namedref{Section}}{\namedref{section}}}
\newcommand{\subsecref}{\@ifstar{\namedref{Subsection}}{\namedref{subsection}}}
\newcommand{\appref}{\@ifstar{\namedref{Appendix}}{\namedref{appendix}}}
\newcommand{\tabref}{\@ifstar{\namedref{Table}}{\namedref{table}}}
\newcommand{\figref}{\@ifstar{\namedref{Figure}}{\namedref{figure}}}
\makeatother

\newcommand{\eqn}[1]{eq.~\eqref{#1}}

\newcommand{\eqns}[2]{eqs.~\eqref{#1} and~\eqref{#2}}

\providecommand{\hypersetup}[1]{}
\providecommand{\texorpdfstring}[2]{#1}

\hypersetup{plainpages=false}
\hypersetup{pdfpagemode=UseNone}
\hypersetup{bookmarksnumbered=true}
\hypersetup{pdfstartview=FitH}
\hypersetup{colorlinks=false}
\hypersetup{citebordercolor={.5 1 .5}}
\hypersetup{urlbordercolor={.5 1 1}}
\hypersetup{linkbordercolor={1 .7 .7}}

\makeatletter
\let\@keywords\@empty
\let\@subject\@empty
\providecommand{\keywords}[1]{\gdef\@keywords{#1}}
\providecommand{\subject}[1]{\gdef\@subject{#1}}
\def\thetitle{\@title}
\def\theauthor{\@author}
\def\thesubject{\@subject}
\def\thedate{\@date}
\def\thekeywords{\@keywords}
\makeatother
\AtBeginDocument{
\hypersetup{pdftitle={\thetitle}}%
\hypersetup{pdfauthor={\theauthor}}%
\hypersetup{pdfsubject={\thesubject}}%
\hypersetup{pdfkeywords={\thekeywords}}%
}

\newif\ifnote 
\notetrue

\allowdisplaybreaks

\newcommand{\ba}  {\begin{array}}
\newcommand{\ea}  {\end{array}}
\newcommand{\bdm} {\begin{displaymath}}
\newcommand{\edm} {\end{displaymath}}

\newcommand{\bea} {\begin{equation}\ba{lcl}}
\newcommand{\eea} {\ea\end{equation}}

\newcommand{\bc}  {\begin{center}}
\newcommand{\ec}  {\end{center}}

\newcommand{\ad} {\mathrm{ad}}

\def\beq{\begin{equation}}
\def\eeq{\end{equation}}

\def\Im{{\rm Im\,}}

\newcommand{\vecb}{\left(\begin{array}{c}}
\newcommand{\vece}{\end{array}\right)}
\newcommand{\ccb}{\left(\begin{array}{cc}}
\newcommand{\cce}{\end{array}\right)}
\newcommand{\cccb}{\left(\begin{array}{ccc}}
\newcommand{\ccce}{\end{array}\right)}
\newcommand{\ccccb}{\left(\begin{array}{cccc}}
\newcommand{\cccce}{\end{array}\right)}
\newcommand{\cccccb}{\left(\begin{array}{ccccc}}
\newcommand{\ccccce}{\end{array}\right)}

\newcommand{\pd}{\partial}

\newcommand{\ve}{\varepsilon}

\newcommand{\ga}{\gamma}

\newcommand{\ep}{\epsilon}

\newcommand{\z}{\zeta}

\newcommand{\om}{\omega}

\newcommand{\te}{\textrm}

\newcommand{\co}{\ , \ \ \ \ \ \ }
\newcommand{\dd}{\mathrm{d}}

\newcommand{\ap}{\alpha'}

\newcommand{\tdz}{\widetilde{z}}

\newcommand{\ZR}{\mathbb R}
\newcommand{\ZC}{\mathbb C}

\newcommand{\ZZ}{\mathbb Z}
\newcommand{\ZQ}{\mathbb Q}
\newcommand{\ZL}{\mathbb L} 

\newcommand{\CH}{\mathcal H}

\newcommand{\CU}{\mathcal U}
\newcommand{\CZ}{\mathcal Z}

\newcommand\mand{\qquad\textrm{and}\qquad}

\newcommand{\nnl}{\nonumber\\}

\DeclareMathOperator{\GL}{\Gamma}
\DeclareMathOperator{\ELi}{ELi}
\DeclareMathOperator{\GGs}{G}
\DeclareMathOperator{\gm}{\gamma}
\DeclareMathOperator{\gmz}{\gamma_0}
\DeclareMathOperator{\zm}{\zeta}
\DeclareMathOperator{\zms}{\zeta^{\shuffle}}
\DeclareMathOperator{\omm}{\omega}
\DeclareMathOperator{\dlog}{\mathrm{dlog}}

\newcommand{\DAlg}{\mathfrak{u}}
\newcommand{\GGn}[1]{\GGs^0_{#1}}
\newcommand{\GG}[1]{\GGs_{#1}}
\newcommand{\GLargz}[2]{\GL\left(\begin{smallmatrix}#1\\#2\end{smallmatrix};z\right)}
\newcommand{\GLarg}[3]{\GL\left(\begin{smallmatrix}#1\\#2\end{smallmatrix};#3\right)}

\newcommand{\fm}{\mathfrak{m}}
\newcommand{\EMZVDatamine}{\texttt{https://tools.aei.mpg.de/emzv}}

\usepackage{amsthm} 
\theoremstyle{plain}

\vspace*{1.8cm}
\title{\textbf{Relations between elliptic multiple zeta values\\and a special derivation algebra}}
\author{Johannes Broedel$^{\te{a}}$, Nils Matthes$^{\te{b}}$, 
Oliver Schlotterer$^{\te{c}}$}
\date{\today}

\begin{document}
\pdfbookmark[1]{Title Page}{title} \thispagestyle{empty}
\begin{flushright}
  \verb!MITP/15-048!
\end{flushright}
\vspace*{0.4cm}
\begin{center}%
  \begingroup\LARGE\bfseries\thetitle\par\endgroup
  \vspace{1.4cm}

\begingroup\large\theauthor\par\endgroup
\vspace{8mm}%

\begingroup\itshape
$^{\te{a}}$Institut f\"ur Theoretische Physik,\\
Eidgen\"ossische Technische Hochschule Z\"urich\\
Wolfgang-Pauli-Strasse 27, 8093 Z\"urich, Switzerland
\par\endgroup
\vspace{4mm}
\begingroup\itshape
$^{\te{b}}$Fachbereich Mathematik, Universit\"at Hamburg,\\
Bundesstra{\ss}e 55, 20146 Hamburg, Germany
\par\endgroup
\vspace{4mm}
\begingroup\itshape
$^{\te{c}}$Max-Planck-Institut f\"ur Gravitationsphysik,\\
Albert-Einstein-Institut\\
Am M\"uhlenberg 1, 14476 Potsdam, Germany
\par\endgroup

\vspace{0.8cm}

\begingroup\ttfamily
jbroedel@ethz.ch, 
nils.matthes@uni-hamburg.de, 
olivers@aei.mpg.de 
\par\endgroup

\vspace{1.2cm}

\bigskip

\textbf{Abstract}\vspace{5mm}

\begin{minipage}{13.4cm}
We investigate relations between elliptic multiple zeta values and describe a
method to derive the number of indecomposable elements of given weight and
length. Our method is based on representing elliptic multiple zeta values as
iterated integrals over Eisenstein series and exploiting the connection with a
special derivation algebra. Its commutator relations give rise to constraints
on the iterated integrals over Eisenstein series relevant for elliptic multiple
zeta values and thereby allow to count the indecomposable representatives.
Conversely, the above connection suggests apparently new relations in the
derivation algebra. Under \EMZVDatamine{} we provide relations for elliptic
multiple zeta values over a wide range of weights and lengths.
\end{minipage}

\vspace*{4cm}

\end{center}

\newpage

\setcounter{tocdepth}{2}
\tableofcontents


\section{Introduction}

While multiple zeta values (MZVs) have been a very active field of research
during the last decades, their elliptic analogues have received more attention
only recently. Pioneered by the work of Enriquez~\cite{EnriquezEMZV},
Levin~\cite{Levin}, Levin and Racinet \cite{LevinRacinet} as well as Brown and
Levin~\cite{BrownLev}, many properties of elliptic multiple zeta values (eMZVs)
have been identified. While mathematically interesting objects in their own
right, eMZVs, the associated elliptic iterated integrals as well as related
objects such as multiple elliptic polylogarithms appear in various contexts in
quantum field theory and string theory. Well-known examples include the
one-loop amplitude in open superstring theory \cite{Broedel:2014vla} as well as
the sunset Feynman integral and its generalizations \cite{Bloch:2013tra,
Adams:2014vja, Bloch:2014qca, Adams:2015gva}. We would also like to mention a
recent application of elliptic functions to the reduction of Feynman integrals
using maximal unitarity cuts \cite{Sogaard:2014jla}. 

Algebraic relations between usual MZVs are well understood based on their
conjectural structure as a Hopf algebra comodule \cite{Brown:2011ik,BrownTate}.
The number of indecomposable MZVs of given weight and depth is expected to be
given by the Broadhurst-Kreimer conjecture \cite{BroadKrei}, which is in line
with Zagier's conjecture \cite{Zagier23} on the counting of MZVs at fixed
weight. A comprehensive collection of relations among MZVs has been made
available in the MZV data mine \cite{Blumlein:2009cf}.

In this article, we investigate relations between eMZVs and classify their
\textit{indecomposable} representatives. Apart from the shuffle relations which
are immediately implied by their definition as iterated integrals, eMZVs are
related by Fay identities. Those identities are the generalization of
partial-fraction identities, which appear in the context of usual MZVs. Both
shuffle and Fay relations preserve the overall modular weight of the integrand
which appears to furnish a natural analogue of the MZVs' transcendental weight.
We will describe a systematic way of exploiting the combination of the two
types of relations.  However, the application of this method to higher weights
and lengths suffers from the proliferating combinatorics of the Fay relations. 

An alternative and computationally more efficient way of studying relations
between eMZVs consists of employing their Fourier expansion in $q= e^{2\pi
i\tau}$, where $\tau$ is the modular parameter of the elliptic curve. The
$q$-derivative of eMZVs is known from ref.~\cite{EnriquezEMZV} in terms of
Eisenstein series and eMZVs of lower length.  Since eMZVs degenerate to MZVs at
the cusp $q\rightarrow 0$ in a manner described in refs.~\cite{EnriquezEMZV,
EnriquezEllAss}, the supplementing boundary value is available as well.  Hence,
the differential equation can be integrated to yield the $q$-expansion of eMZVs
recursively to -- in principle -- arbitrarily high order. Once the
$q$-expansion of an eMZV is available up to a certain power in $q$, finding
 relations between eMZVs valid up to this particular power
and identifying indecomposable representatives amounts to solving a linear
system. 

Clearly, the agreement of the respective $q$-expansions up to a certain power
in $q$ is necessary but not sufficient for the validity of eMZV relations.
Nevertheless, the method of $q$-expansions allows to confirm that indeed, Fay
and shuffle identities comprise the entirety of eMZV relations for a variety of
combinations of weights and lengths.  Accordingly, this leads us to conjecture
that \textit{all} available relations between eMZVs are implied by Fay and
shuffle identities. At lengths and weights beyond the reach of our current
computer implementation of Fay and shuffle identities, the comparison of
$q$-expansions gives rise to conjectural relations which nicely tie in with the
algebraic considerations to be described next.

In order to overcome the shortcomings of comparing $q$-expansions of eMZVs, one
uses their differential equation in $\tau$ to write eMZVs as linear
combinations of iterated integrals over Eisenstein series or \textit{iterated
Eisenstein integrals} for short.  Contrary to the definition of eMZVs, where
the iterated integration is performed over coordinates of the elliptic curve,
the integration in their representation via Eisenstein series is performed over
the modular parameter of the elliptic curve. Similar iterated integrals, some
of them involving more general modular forms, have been studied by Manin in
ref.~\cite{ManinIterMod}. Those integrals have been revisited by Brown
\cite{Brown:mmv} recently, who used them to define \textit{multiple modular
values}. A new feature of Brown's approach to iterated integrals of modular
forms is that it allows also for an integration along paths which connect two
cusps. Among other things, multiple modular values provide a conceptual
explanation of the relationship between double zeta values and cusp forms
\cite{GKZ}.

The representation of eMZVs as iterated Eisenstein integrals is particularly
convenient because the latter are believed to be linearly independent over the
complex numbers. This has been tested up to the lengths and weights considered
in this paper, but it remains a working hypothesis for several constructions in
this work\footnote{It is expected that $\mathbb C$-linear independence of
  iterated Eisenstein integrals holds in full generality, and it can presumably
  be shown using the techniques introduced in \cite{Hain2014}. We do not
attempt a proof in this paper and defer this problem to future work
\cite{Matthes:thesis}.}. In particular, an analogue of the Fay relations is not
known to hold for iterated Eisenstein integrals. This led to the first
expectation that one can find the number of indecomposable eMZVs by enumerating
all shuffle-independent iterated Eisenstein integrals.

While this idea indeed yields the correct number of indecomposable eMZVs of low
length and weight, there is another effect appearing at higher weight: in the
rewriting of \textit{any} eMZV certain shuffle-independent iterated Eisenstein
integrals occur in rigid linear combinations only. In other words, not all
iterated Eisenstein integrals can be expressed in terms of eMZVs.  The above
rigid linear combinations in turn are implied by relations well known from a
special algebra of derivations $\DAlg$ \cite{KZB,Hain,Pollack,Brown:1504}. The
patterns we find from investigating various eMZVs exactly match the available
data about the derivation algebra in refs.~\cite{Pollack, LNT}. Consequently, we
turn this into a method to infer the number of indecomposable eMZVs at given
weight and length.  The results of this method agree perfectly with the data
obtained from either shuffle and Fay relations. In addition, complete knowledge
of relations in the derivation algebra leads to upper bounds on the number of
indecomposable eMZVs. Those upper bounds complement the lower bounds obtained
from comparing $q$-expansions. The study of $q$-expansions allows to enumerate
indecomposable eMZVs of given weight and length without assuming linear
independence of different iterated Eisenstein integrals, which will be
discussed below. 

The algebra of derivations $\DAlg$ on one side and eMZVs on the other side are
linked by a differential equation for the elliptic KZB associator~\cite{KZB,
Hain} derived by Enriquez~\cite{EnriquezEllAss}. This differential equation
implies upper bounds on the number of indecomposable eMZVs.  Assuming these
upper bounds to be attained, one can extend the knowledge about the derivation
algebra $\DAlg$ substantially: we identify numerous apparently new relations up
to and including depth five. Moreover, there is no conceptual bottleneck in
extending the analysis to arbitrary weight and depth.

The link between eMZVs and the derivation algebra $\DAlg$ becomes particularly
clear upon mapping iterated Eisenstein integrals onto words composed from
non-commutative letters $g$.  The viability of the bookkeeping framework
introduced in section \ref{sec:gamma} relies on the linear independence of
iterated Eisenstein integrals, which is not proved at this point. On the other
hand, since one can always check linear independence for a finite number of
iterated Eisenstein integrals directly, the empirical results made available on
our eMZV webpage do not depend on the general linear independence
statement.

Moreover, the rewriting of eMZVs as linear combinations of iterated Eisenstein
integrals and into letters $g$ lateron bears similarities to a process, which
appeared already in the context of usual MZVs.  Namely, employing a conjectural
isomorphism $\phi$, MZVs can be rewritten in terms of an alphabet of
non-commutative letters $f$ (cf.~ref.~\cite{Brown:2011ik}). However, the
construction of the map $\phi$ is highly elaborate, as it requires the motivic
coaction and also depends on the choice of an algebra basis for (motivic) MZVs.
On the contrary, the rewriting of eMZVs in terms of iterated Eisenstein
integrals is completely canonical and straightforward from the differential
equation for eMZVs.  On the other hand, while the number of indecomposable MZVs
at given weight is determined by counting {\em all} shuffle-independent words
in $f$, the corresponding problem for eMZVs requires the consideration of the
non-trivial relations in the derivation algebra $\DAlg$ in addition.

In summary, the results in this work are fourfold:
\begin{itemize}
\item an explicit basis of irreducible eMZVs of certain weights and lengths,
  together with expressions of eMZVs in that basis collected on a web page
  \EMZVDatamine{}
\item the observation that Fay, shuffle and reflection relations generate all
  identities between eMZVs up to the weights and lengths considered.
\item a general method for counting irreducible eMZVs based on iterated
  Eisenstein integrals and the special derivation algebra, for which we need to
  assume linear independence of iterated Eisenstein integrals.
\item several explicit new relations in the special derivation algebra $\DAlg$,
  which are collected also at \EMZVDatamine{}
\end{itemize}
The article is organized as follows: \secref{sec:eMZVrel} starts with a small
review of eMZVs and sets the stage for combining Fay and shuffle relations with
the $q$-expansion, resulting in an ``observational'' set of indecomposable
eMZVs. \secref*{sec:MZV} is devoted to a brief recapitulation of structures
present for usual MZVs with particular emphasis on their rewriting in terms of
non-commutative words. In \secref{sec:gamma} we set up the translation of eMZVs
into iterated Eisenstein integrals, investigate their properties and connect
the bookkeeping of indecomposable eMZVs with the algebra of derivations
$\DAlg$. In \subsecref{sec:simple} we describe a modified version of iterated
Eisenstein integrals suitable in particular for the description of eMZVs.
Several appendices are complementary to the discussion in the main text. In
particular, some relations between elements of the derivation algebra are collected in
\appref{sec:explrel}.


\section{Relations between elliptic multiple zeta values}
\label{sec:eMZVrel}

After recalling the definition of eMZVs, we are going to explore the
implications of Fay and shuffle relations as well as the method of
$q$-expansions.  In addition, we will describe how usual MZVs defined via 
\begin{equation}
\zm_{n_1,n_2,\ldots,n_r} \equiv  \sum_{0<k_1<k_2<\ldots<k_r}k_1^{-n_1}k_2^{-n_2}\ldots k_r^{-n_r}\co  n_r\geq2 
\label{mzvsum}
\end{equation}
arise as constant terms of eMZVs. 


\subsection{Prerequisites and definitions}
\label{ssec:prerequisites}

In this subsection we will briefly review elliptic iterated integrals and define
eMZVs. An elaborate introduction from a string theorist's point of view is
available in ref.~\cite{Broedel:2014vla}.  To get started, let us consider
iterated integrals on the punctured elliptic curve $E_\tau^\times$, which is
$E_\tau\equiv \ZC/(\ZZ+\ZZ\tau)$ with the origin removed and $\Im(\tau)>0$.  We
will frequently refer to the modular parameter $\tau$ by its exponentiated
version
\begin{equation}
  q\equiv e^{2\pi i\tau},\quad\text{such that}\quad 2\pi i\frac{\dd}{\dd \tau}=-4\pi^2 q \frac{\dd}{\dd q}=-4 \pi^2 \frac{\dd}{\dlog q} \ .
  \label{eqn:qtau}
\end{equation}
Functions $A$ of the modular parameter will be denoted by either $A(\tau)$
or $A(q)$.


\paragraph{Weighting functions.}
A natural collection of weighting functions for the iterated integration to be
defined below is provided by the Eisenstein-Kronecker series $F(z,\alpha,\tau)$
\cite{Kronecker,BrownLev},
\begin{equation}
F(z,\alpha,\tau) \equiv 
\frac{\theta_1'(0,\tau)\theta_1(z+\alpha,\tau)}{\theta_1(z,\tau)\theta_1(\alpha,\tau)} \ ,
\label{alt1}
\end{equation}
where $\theta_1$ is the odd Jacobi theta function and the tick denotes a derivative
with respect to the first argument. The definition \eqn{alt1} immediately
yields $F(z,\alpha,\tau)=F(z+1,\alpha,\tau)$, and supplementing an additional,
non-holomorphic factor lifts the quasi-periodicity of the Eisenstein-Kronecker
series with respect to $z \mapsto z+\tau$ to an honest double-periodicity. The
resulting function $\Omega(z,\alpha,\tau)$ on an elliptic curve serves as a
generating series for the weighting functions $f^{(n)}(z,\tau)$ in eMZVs: 
\begin{equation}
\Omega(z,\alpha,\tau)\equiv 
\exp \bigg( 2\pi i \alpha
\frac{\Im(z)}{\Im(\tau)} \bigg) F(z,\alpha,\tau)=
\sum_{n=0}^{\infty}f^{(n)}(z,\tau)\alpha^{n-1}\,.
\label{alt5}
\end{equation}
The functions $f^{(n)}$ are doubly periodic and alternate in their parity,
\begin{equation}
 f^{(n)}(z+1,\tau)=
 f^{(n)}(z+\tau,\tau)=
 f^{(n)}(z,\tau)\, ,
\qquad
  f^{(n)}(-z,\tau)=(-1)^{n}f^{(n)}(z,\tau)\,.
  \label{eqn:fparity}
\end{equation}
Their simplest instances read
\begin{equation}
 f^{(0)}(z,\tau)=1 \, ,
\qquad
  f^{(1)}(z,\tau)= \frac{ \theta_1'(z,\tau)}{\theta_1(z,\tau)} + 2\pi i \frac{ \Im(z) }{\Im (\tau)}
  \, ,
  \label{eqn:expl}
\end{equation}
and $f^{(1)}$ is in fact the only weighting function with a simple pole on the
lattice $\ZZ+\ZZ\tau$ including the origin. The remaining $f^{(n)}$ with $n\neq
1$ are non-singular on the entire elliptic curve. As elaborated in
\cite{BrownLev} and section 3 of \cite{Broedel:2014vla}, the weighting
functions  $f^{(n)}$ can be expressed in terms of Eisenstein functions and
series the latter of which will play a central r\^ole in the sections below.


\paragraph{Elliptic iterated integrals and eMZVs.} Even though the functions
$f^{(n)}$ are defined for arbitrary complex arguments $z$ and suitable for
integrations along both homology cycles of the elliptic curve, we will restrict
our subsequent discussion to real arguments. This is sufficient for studying
eMZVs as iterated integrals over the interval $[0,1]$ on the real axis and
avoids the necessity for homotopy-invariant completions of the
integrands\footnote{A generating series for  homotopy-invariant iterated
  integrals is given in ref.~\cite{BrownLev}, in which the differential forms
  $f^{(n)}(z,\tau)\dd z$ are accompanied by $\nu \equiv 2\pi i\,\frac{\dd
  \Im(z)}{\Im(\tau)}$. While any integral based upon a sequence of $\nu$ and
  $\dd z$ has a unique homotopy-invariant uplift via admixtures of
  $f^{(n>0)}(z,\tau)\dd z$, iterated integrals of $f^{(n)}(z,\tau)\dd z$ allow for
  multiple homotopy-invariant completions via $\nu$. A thorough discussion of
  the issue is provided in ref.~\cite{Broedel:2014vla}.}.  Hence, any
  integration variable and first argument of $f^{(n)}(z,\tau)$ is understood to be
  real.

Employing the functions $f^{(n)}$, iterated integrals on the elliptic curve
$E_\tau^\times$ are defined via
\begin{equation}
  \GLargz{n_1 &n_2 &\ldots &n_r}{a_1 &a_2 &\ldots &a_r} \equiv
  \int^z_0 \dd t \, f^{(n_1)}(t-a_1) \,
  \GLarg{n_2 &\ldots &n_r}{a_2 &\ldots &a_r}{t},
  \label{eqn:defGell}
\end{equation}
where the recursion starts with $\GL(;z) \equiv 1$. 
The elliptic iterated integral in \eqn{eqn:defGell} is
said to have \textit{weight} $w=\sum_{i=1}^r n_i$, and the number $r$ of
integrations will be referred to as its \textit{length}.
Beginning with the above
equation, we will usually suppress the second argument $\tau$ for the 
weighting functions $f^{(n)}$ and the elliptic iterated integrals $\GL$.

Definition (\ref{eqn:defGell}) implies a shuffle relation with respect to the
combined letters $A_i \equiv \begin{smallmatrix} n_i  \\ a_i \end{smallmatrix}$
  describing the weighting functions $f^{(n_i)}(z-a_i)$,
\begin{equation}
\GL(A_1,A_2,\ldots,A_r;z) \GL(B_1,B_2,\ldots,B_q;z)  = \GL\big(( A_1,A_2,\ldots,A_r) 
\shuffle (B_1,B_2,\ldots,B_q);z\big)\,.
\label{gl2}
\end{equation}
Another obvious property of elliptic iterated integrals is the
reflection identity due to \eqn{eqn:fparity}
\begin{equation}
      \GLargz{n_1 &n_2 &\ldots &n_r}{a_1 &a_2 &\ldots &a_r}
      =(-1)^{n_1+n_2+\ldots+n_r}
        \GLargz{n_r &\ldots &n_2 &n_1}{z-a_r &\ldots &z-a_2 &z-a_1}\,.
     \label{eqn:reflection}
\end{equation}
Finally, if all the labels $a_i$ vanish -- which, because of the periodicity of
$f^{(n)}$ is equivalent to all labels $a_i$ being integer -- we will often use the
notation
\begin{equation}
  \GL(n_1,n_2,\ldots,n_r;z) \equiv \GLargz{n_1 &n_2 &\ldots &n_r}{0 &0 &\ldots &0}\,.
\label{eqn:vana}
\end{equation}

Evaluating the elliptic iterated integrals in \eqn{eqn:vana} at $z=1$ gives
rise to \textit{elliptic multiple zeta values} or \textit{eMZV}s for
short~\cite{EnriquezEMZV}:
\begin{align}
\omm(n_1,n_2,\ldots,n_r) &\equiv \! \! \! \! \! \! \int \limits_{0\leq z_i \leq z_{i+1} \leq 1}  \! \! \! \! \! \!
f^{(n_1)}(z_1) \dd z_1\, f^{(n_2)}(z_2)\dd z_2\, \ldots f^{(n_r)}(z_r) \dd z_r 
\label{l2.7} \\
&= 
\GL(n_r  ,\ldots ,n_2 ,n_1;1) \,,\notag
\end{align}
where $\ell_\om=r$ is referred to as the \textit{length} while
$w_\om=\sum_{i=1}^r n_i$ is called the \textit{weight} of an eMZV. The
subscript $\omm$ refers to the current $\omm$-representation of eMZVs in
\eqn{l2.7}, which has a different notion of weight and length compared to the
iterated Eisenstein integrals to be defined below in \secref{sec:gamma}.

Being defined on an elliptic curve, eMZVs depend on its modular parameter
$\tau$ and furnish a natural genus-one generalization of standard MZVs, which
are to be reviewed in \secref{sec:MZV}.  

The shuffle relation \eqn{gl2} straightforwardly
carries over to eMZVs,
\begin{equation}
\omm(n_1,n_2,\ldots,n_r) \omm(k_1,k_2,\ldots,k_s) = \omega\big( (n_1,n_2,\ldots,n_r) \shuffle (k_1,k_2,\ldots,k_s) 
\big)\,,
\label{l2.7sh}
\end{equation}
whereas the parity property \eqn{eqn:fparity} of the
weighting functions $f^{(n)}$ implies the reflection identity
\begin{equation}
\omm(n_1,n_{2} ,\ldots ,n_{r-1} ,n_r)=
     (-1)^{n_1+n_2+\ldots+n_r}\omm(n_r,n_{r-1} ,\ldots ,n_2 ,n_1)\,.
\label{eqn:reflect}
\end{equation}
Note that this implies the vanishing of odd-weight eMZVs with reversal-symmetric labels:
\begin{align}
  \omm(n_1,n_2,\ldots,n_r)&=0 \, , \qquad\text{if}\;(n_1,n_2,\ldots,n_r)=(n_r,\ldots,n_2,n_1)\ \text{and}\ \sum_{i=1}^r n_i\ \text{odd}\,.
\end{align}
Although suppressed in our notation, every eMZV is still a function of the
modular parameter $\tau$ and inherits a Fourier expansion in $q$ from the restriction of $f^{(n)}$
to real arguments.
\begin{equation}
  \omm(n_1,\ldots,n_r)=\om_0(n_1,\ldots,n_r)+\sum_{k=1}^{\infty}c_k(n_1,\ldots,n_r) q^k \ .
  \label{eqn:qexp}
\end{equation}
The $\tau$-independent quantity $\omega_0$ in \eqn{eqn:qexp} is called the
\textit{constant term} of $\omega$ and will be shown to consist of MZVs and integer powers of
$2\pi i$ in the next subsection. We will refer to eMZVs for which
$c_k(n_1,\ldots,n_r)=0$ for all $k \in \mathbb N^+$ as \textit{constant}.


\paragraph{Regularization.} While the functions $f^{(n)}(z)$ are smooth for $n \neq 1$, the function $f^{(1)}(z)$ in \eqn{eqn:expl}
diverges as $\frac{1}{z}$ and $\frac{1}{z-1}$ for $z \to 0$ and $z \to 1$,
respectively. Hence, eMZVs $\omega(n_1,\ldots,n_r)$ with $n_1=1$ or $n_r=1$ are
a priori divergent, and require a regularization process similar to shuffle
regularization for MZVs~\cite{EnriquezEMZV} (cf.~also~\cite{Matthes:thesis}). A
natural choice at genus one is to modify the integration region in
\eqn{eqn:defGell} by a small $\varepsilon>0$,
\begin{equation}
\int \limits_{\varepsilon \leq z_i \leq z_{i+1} \leq z-\varepsilon}  \! \! \! \! \! \!
f^{(n_1)}(z_1-a_1) \dd z_1\, f^{(n_2)}(z_2-a_2)\dd z_2\, \ldots f^{(n_r)}(z_r-a_r) \dd z_r \ ,
\label{eqn:shufreg}
\end{equation}
and to expand the integral as a polynomial in $\log(-2\pi i \varepsilon)$.
Hereby the branch of the logarithm is chosen such that $\log(-i)=-\frac{\pi
i}{2}$.  The regularized value of \eqn{eqn:shufreg} is then defined to be the
constant term in the $\ve$-expansion. The factor $-2\pi i$ in the expansion
parameter $\log(-2\pi i \varepsilon)$ ensures that the limit $\tau \to i
\infty$ does not introduce any logarithms, and that eMZVs degenerate to MZVs
upon setting $z=1$ in \eqn{eqn:shufreg}. For later reference, we will call
eMZVs of the form $\omega(1,n_2,\ldots)$ or $\omega(\ldots,n_{r-1},1)$
\emph{divergent}.

For the enumeration of eMZVs, we have employed an infinite alphabet, consisting
of the non-negative integers $0,1,2,\ldots$ \eqn{l2.7}. There is another way of
carrying out this enumeration, which uses a two-letter alphabet instead
\cite{BrownLev}. The two-letter alphabet descends from a construction of eMZVs
via \textit{homotopy-invariant} iterated integrals.  Since every eMZV in the
infinite alphabet can be rewritten as an eMZV in the two-letter alphabet and
vice-versa, one does not lose information by choosing to work with one alphabet
or the other.


\subsection{Fay and shuffle relations}
\label{ssec:rels}

In this subsection, we analyze relations among eMZVs defined in \eqn{l2.7} and
gather examples of {\em indecomposable} eMZVs.  A set of
\textit{indecomposable} eMZVs of weight $w_\om$ and length $\ell_\om$ is a
\textit{minimal} set of eMZVs such that any other eMZV of the same weight and
length can be expressed as a linear combination of elements from this set and
\begin{itemize}
  \item products of eMZVs with strictly positive weights,
  \item eMZVs of lengths smaller than $\ell_\om$ or weight lower than $w_\om$, 
\end{itemize}
where the coefficients comprise MZVs (including rational numbers) and integer
powers of $2\pi i$.  After exploring the consequences of shuffle and reflection
identities \eqns{l2.7sh}{eqn:reflect}, Fay identities are discussed as
a genus-one analogue of the partial-fraction identities among products of
$(z-a)^{-1}$, which arise from the differential forms $ \dlog(z-a)$. The weight of eMZVs
is preserved under all these identities whereas the length obviously varies in
Fay and shuffle relations. 
In contradistinction to usual MZVs, the availability of $f^{(0)} \equiv 1$ 
as a weighting function yields an infinite number of eMZVs for a certain weight, 
so the counting of indecomposable eMZVs must be performed at fixed length and weight.


\paragraph{Examples of constant eMZVs.} The simplest examples of the eMZVs
defined in \eqn{l2.7} are of length one:
\begin{equation}
\omm(n_1) =\left\{ \begin{array}{cl}-2\zm_{n_1} &: \ n_1 \ \te{even} \\ 0 &: \ n_1 \ \te{odd} \end{array} \right.\,.
\label{eqn:ellone}
\end{equation}
The underlying single integration over the interval $[0,1]$ picks up the
constant term in the \mbox{$q$-expansion} of $f^{(n)}$ (see section 3.3 of
ref.~\cite{Broedel:2014vla}) and yields the constants in \eqn{eqn:ellone} with 
regularized value $\zeta_0=-\frac{1}{2}$ in $\omm(0)=1$.

Another distinction between even and odd labels $n_i$ occurs at length
$\ell_\om=2$. The union of shuffle and reflection identities \eqns{l2.7sh}{eqn:reflect} 
contains more independent relations for even total weight than for odd weight, 
and the eMZVs are then completely determined by \eqn{eqn:ellone}:
\begin{equation}
\omm(n_1,n_2) \, \Big|_{n_1+n_2 \ \te{even}} = \left\{ \begin{array}{cl}  2\zm_{n_1}\zm_{n_2} &: \ n_1,n_2 \ \te{even} \\ 0 &: \ n_1,n_2 \ \te{odd} \end{array} \right. .
  \label{eqn:constEMZV}
\end{equation}
For eMZVs of odd total weight, on the other hand, shuffle and reflection
relations at length two coincide, and $\omm(n_1,n_2)$ are no longer bound to be
constant.  This correlation between $(-1)^{w_\omega}$ and the length will be
turned into a general rule in the next paragraph.
In addition, there are also constant eMZVs, which make their appearance only at
sufficiently high length. For example, $\zm_3$ is identified in
\eqn{eqn:zeta3example} to be an eMZV of weight $3$ and length $4$. One can show
that all constant eMZVs evaluate to products of MZVs and integer powers of $2\pi
i$, see Proposition 5.3 of ref.~\cite{EnriquezEMZV}.


\paragraph{Interesting and boring eMZVs.} The lack of a $\tau$-dependence for
eMZVs $\omm(n_1,n_2)$ of even weight can be viewed as the analogue of the
vanishing of $\omm(n_1)$ for odd weight as observed in \eqn{eqn:ellone}. The
general pattern is as follows: whenever weight and length of an eMZV have the
same parity (i.e.~$(-1)^{w_\om}=(-1)^{\ell_\om}$), shuffle and reflection
identities \eqns{l2.7sh}{eqn:reflect} allow to determine this eMZV in terms of
eMZVs of lower length.  Novel indecomposable eMZVs can only occur for opposite
parity $(-1)^{w_\om}=-(-1)^{\ell_\om}$ such as the odd-weight
$\omega(n_1,n_2)$.  Accordingly, an eMZV $\omm(n_1,n_2,\ldots,n_r)$ is called
\textit{interesting}, if the combination $w_\om+\ell_\om$ of weight and length
is odd, otherwise we refer to it as \textit{boring}.

Boring eMZVs at length $\ell_\omega=3$ can arise from four different choices of
even and odd labels. For those, the shuffle identities \eqn{l2.7sh} allow to reduce
them to interesting eMZVs at length two. Explicitly, we have
\begin{align}
\omm(o_1,o_2,o_3)&=0\nnl
\omm(e_1,e_2,o_3)&=-\zm_{e_1} \omm(e_2,o_3) \nnl
\omm(e_1,o_2,e_3)&=-\zm_{e_1} \omm(o_2,e_3) - \zm_{e_3} \omm(e_1,o_2) \nnl
\omm(o_1,e_2,e_3)&=-\zm_{e_3} \omm(o_1,e_2)\,,
\label{mine3}
\end{align}
where $e_i$ and $o_i$ refer to even and odd labels, respectively. Similarly,
boring eMZVs at length $\ell_\omega=4$ come in the following
(reflection-independent) classes:
\begin{align}
\omm(o_1,o_2,o_3,o_4) &=0
\notag
\\
\omm(e_1,e_2,e_3,e_4) &= -2 \zm_{e_1} \zm_{e_2} \zm_{e_3} \zm_{e_4} - \zm_{e_4} \omm(e_1,e_2,e_3) - \zm_{e_1} \omm(e_2,e_3,e_4) \notag
\\
\omm(o_1,o_2,e_3,e_4) &= - \zm_{e_4} \omm(o_1,o_2,e_3) \label{sh42} 
\\
\omm(o_1,e_2,o_3,e_4) &= \tfrac{1}{2} \omm(o_1,e_2) \omm(o_3,e_4) - \zm_{e_4} \omm(o_1,e_2,o_3) \notag \\
\omm(o_1,e_2,e_3,o_4) &= \tfrac{1}{2} \omm(o_1,e_2) \omm(e_3,o_4) \notag\\
\omm(e_1,o_2,o_3,e_4) &= \tfrac{1}{2} \omm(e_1,o_2) \omm(o_3,e_4) - \zm_{e_1} \omm(o_2,o_3,e_4) - \zm_{e_4} \omm(e_1,o_2,o_3)\,.
\notag
\end{align}
Although becoming more involved for higher length, the distinction of cases as
well as the decomposition of boring eMZVs can be cast into a nice form, as is
explained in \appref{app:boringeMZVs}.  Below, however, we will be concerned
with interesting eMZVs mostly. Note that the vanishing of eMZVs with only odd
entries is true at all lengths,
\begin{equation}
\omm(o_1,o_2,\ldots,o_r)=0 \ .
\label{eqn:lm}
\end{equation}
%

\paragraph{Fay relations among $\boldsymbol{f^{(n)}}$ and elliptic iterated
integrals.} While reflection and shuffle identities preserve the partition of
the modular weight among the integrated $f^{(n_i)}$, so-called Fay relations
mix eMZVs involving different values of $n_i$. They can be traced back to the
Fay identity of their generating series \eqn{alt5} \cite{BrownLev}
\begin{align}
\Omega(z_1,\alpha_1,\tau)\Omega(z_2,\alpha_2,\tau)&=\Omega(z_1,\alpha_1+\alpha_2,\tau)\Omega(z_2-z_1,\alpha_2,\tau) \notag
				\\
&\qquad+\Omega(z_2,\alpha_1+\alpha_2,\tau)\Omega(z_1-z_2,\alpha_1,\tau) \ ,\label{alt4}
\end{align}
which is valid for any complex $z_{1},z_{2}$ and follows from the Fay trisecant
equation \cite{mumford1984tata}. Relations among $f^{(n)}$ can be read off from
\eqn{alt4} by isolating monomials in $\alpha_1, \alpha_2$
\cite{Broedel:2014vla}
\begin{align}
f^{(n_1)}(t-x) f^{(n_2)}(t)
 &=  - (-1)^{n_1} f^{(n_1+n_2)}(x) + \sum_{j=0}^{n_2} { n_1 - 1 + j \choose j} f^{(n_2-j)}(x) f^{(n_1+j)}(t-x) \notag  \\
 & \ \ \ \ \ + \sum_{j=0}^{n_1} {n_2-1+j \choose j} (-1)^{n_1+j} f^{(n_1-j)}(x) f^{(n_2+j)}(t) \ .
\label{kron1.33b}
\end{align}
The simplest instance of these Fay relations can be viewed as a genus-one
counterpart of partial-fraction relations such as $\frac{1}{tx}= \frac{1}{x(t-x)}
+ \frac{1}{t(x-t)}$:
\begin{equation}
f^{(1)}(t-x) f^{(1)}(t) = f^{(1)}(t-x) f^{(1)}(x) - f^{(1)}(t) f^{(1)}(x) + 
f^{(2)}(t) + f^{(2)}(x) + f^{(2)}(t-x)  \ .
\label{kron1.34}
\end{equation}
The Fay relations \eqn{kron1.33b} are a very powerful tool for rearranging the
elliptic iterated integrals in \eqn{eqn:defGell}. Together with the derivatives
of $\Gamma$ with respect to their argument $z$ and labels $a_i$
\cite{Broedel:2014vla}, they allow for example to recursively remove any
appearance of $a_i=z$ in the label of an iterated integral, e.g.
\begin{align}
\GLarg{n_1 &n_2 &\ldots &n_r}{z &0 &\ldots &0}{z} &= 
 (-1)^{r} \zm_{r} \prod_{j=1}^r \delta_{n_j,1} - 
  (-1)^{n_1} \GLarg{n_1+n_2 &0 &n_3 &\ldots &n_r}{0 &0 &0&\ldots &0}{z} \nnl
&+\sum_{j=0}^{n_1} (-1)^{n_1+j} {n_2-1+j \choose j}  
  \GLarg{n_1-j &n_2+j &n_3 &\ldots &n_r}{0 &0 &0 &\ldots &0}{z}   \nnl
&+\sum_{j=0}^{n_2} {n_1-1+j \choose j} \int^z_0 \dd t \, f^{(n_2-j)}(t) 
  \GLarg{n_1+j &n_3 &\ldots &n_r}{t &0 &\ldots &0}{t} \ , \label{extra21}
  \end{align}
see \appref{app:moreFay} for a generalization to multiple appearances of
$a_i=z$. The zeta value $\zm_{r}$ in the first line of \eqn{extra21} stems from
the limit $z\rightarrow 0$ of the left hand side for which $f^{(1)}(z)$ can be
approximated by $\frac{1}{z}$ \cite{Broedel:2014vla}. Note that the
Kronecker-deltas $\delta_{n_j,1}$ ensure that the notions of weights for MZVs
and elliptic iterated integrals are compatible in \eqn{extra21}.
%

\paragraph{Fay relations among eMZVs.} A rich class of eMZV relations can be
inferred from the limit $z\rightarrow 1$ of \eqn{extra21}. On the left hand
side, periodicity of $f^{(n)}$ w.r.t.~$z\rightarrow z+1$ leads to
\begin{equation}
\lim_{z\rightarrow 1}\GLarg{n_1 &n_2 &\ldots &n_r}{z &0 &\ldots &0}{z} = \omm(n_r,\ldots,n_2,n_1) \co n_1\neq1 \ \te{or} \ n_2\neq 1 \ ,
\label{lhs}
\end{equation}
where cases with $n_1=n_2=1$ require an additional treatment of the poles of
the associated $f^{(1)}$ and are therefore excluded\footnote{For cases with
  $n_1=1$ and $n_2\neq 1$, we could not prove the general absence of extra
  contributions from the poles of $f^{(1)}$. However, the validity of \eqn{lhs}
in these cases has been thoroughly tested to lengths $r\leq 6$ using the
methods in section \ref{ssec:constq}. Hence, \eqn{lhs} at $n_1=1$ and $n_2\neq
1$ with general $r$ remains a well-tested conjecture.}. By \eqn{l2.7}, the
elliptic iterated integrals on the right hand side reduce to eMZVs under
$z\rightarrow 1$ once the recursion \eqn{extra21} has been applied iteratively
to remove any appearance of the argument from the labels. At length two, the
resulting eMZV relation is
\begin{align}
\omm( n_2,n_1)&=  - (-1)^{n_1} \omm( 0,n_1 + n_2) + \sum_{j=0}^{n_2} {  n_1 - 1 + j\choose j} (-1)^{n_1 + j} \omm( n_1 + j,n_2 - j)\notag \\
& \ \  \ \  \ \  \ \  \ + \sum_{j=0}^{n_1} { n_2 - 1 + j \choose j} (-1)^{n_1 + j} \omm( n_2 + j,n_1 - j) \co n_1\neq1 \ \te{or} \ n_2\neq 1 \ ,
\label{sumexp1}
\end{align}
and length three requires two applications of the recursion in \eqn{extra21}:
\begin{align}
&\omm(n_3 ,n_2 ,n_1) =   
\zeta_2 \sum_{ j=0}^{n_2} \delta_{n_3,1} \delta_{n_1+j,1} {n_1 - 1 + j\choose j}  \omm(n_2 - j) \notag \\
& - (-1)^{n_1} \omm( n_3,0,n_1 + n_2) +\sum_{j=0}^{n_1}  (-1)^{n_1 + j} {n_2 - 1 + j\choose j} \omm(n_3, n_2 + j, n_1 - j)  
\label{sumexp2} \\
    & + \sum_{j=0}^{n_2}  {n_1 - 1 + j\choose   j} \sum_{k=0}^{n_3} (-1)^{n_1 + j + k}{n_1 + j - 1 + k \choose k} \omm(n_1 + j + k ,  n_3 - k, n_2 - j   ) \notag \\
      & + \sum_{j=0}^{n_2} {n_1 - 1 + j\choose    j} \sum_{k=0}^{n_1+j} (-1)^{n_1 + j + k} {n_3 - 1 + k\choose k} \omm(n_3 + k, 
      n_1 + j - k,   n_2 - j) \notag \\
& -  \sum_{j=0}^{n_2} (-1)^{n_1 + j} {n_1 - 1 + j\choose j} \omm(0, n_1 + n_3 + j, n_2 - j)  
  \co n_1\neq1 \ \te{or} \ n_2\neq 1 \ .
    \notag
\end{align}
It is straightforward to derive higher-length relations (involving any
$\zm_{r}$ with $2\leq r \leq \ell_\omega-1$) from further iterations of
\eqn{extra21} in the limit $z\rightarrow 1$. The exclusion of $n_1=n_2=1$
suppresses $\zm_{2}$ in \eqn{sumexp1} and $\zm_{3}$ in \eqn{sumexp2}, and, more
generally, the appearance of $\zm_{r}$ is relegated to eMZV relations of length
$r+1$. By the relations derived from $\GLarg{n_1 &n_2 &\ldots &n_{k} &n_{k+1}
&\ldots &n_r}{z &z &\ldots &z &0 &\ldots  &0}{z}$ in \appref{app:moreFay}, 
analogous statements apply to generic MZVs, and any MZV will appear in the rewriting
of some $\GLarg{n_1 &n_2 &\ldots &n_r}{a_1 &a_2&\ldots   &a_r}{z}$ with
appropriate combinations of $a_j \in \{0,z\}$. 
%

\paragraph{Combining shuffle and Fay relations.} Shuffle relations reduce
boring eMZVs to interesting eMZVs of lower length, see for instance
\eqns{eqn:constEMZV}{mine3}. At first glance, this appears to attribute more
significance to Fay relations among interesting eMZVs, e.g.~\eqn{sumexp1} at
odd $n_1+n_2$ and \eqn{sumexp2} at even $n_1+n_2+n_3$.  The former yields
length-two relations such as
\begin{equation}
\omm(0,5)=\om(2,3) \co \omm(3, 4) = -2 \omm(0, 7) + \omm(2, 5) \ ,
\label{fayl2}
\end{equation}
which by themselves leave $1+\lfloor \frac{1}{3}(n_1+n_2) \rfloor$ eMZVs at
length $\ell_\om=2$ and weight $w_\om=n_1+n_2$ independent \cite{matthes:edzv}.
However, Fay relations \eqn{sumexp2} among boring eMZVs at length three turn
out to contain additional information about interesting $\omm(n_1,n_2)$. For
example, writing \eqn{sumexp2} with $(n_1,n_2,n_3)=(1,0,2)$,
\begin{equation}
\omm(0, 3, 0) - \omm(1, 2, 
   0) - \omm(2, 0, 1) + \omm(3, 0, 0)=0 \ ,
\label{fayl3}
\end{equation}
followed by a shuffle-reduction of the boring eMZVs via \eqn{mine3} yields the
length-two relation
\begin{equation}
\omm(1, 2) = 2 \zm_{2} \omm(0, 1) - \omm(0, 3) \ .
\label{fayfay3}
\end{equation}
This relation would be inaccessible from Fay relations at length two and
identifies $\omega(0,3)$ to be the unique indecomposable $\omega(n_1,n_2)$ at
weight three, which is short of the above $1+\lfloor \frac{1}{3}(n_1+n_2)
\rfloor$ counting. Hence -- when combined with shuffle-relations -- Fay
relations among boring eMZVs at length $\ell_\om+1$ provide more information
than their counterparts among interesting eMZVs at length $\ell_\om$. The need
for Fay relations at length $\ell_\om+1$ to classify indecomposable eMZVs at
length $\ell_\om$ is reminiscent of double-shuffle relations among MZVs. For
example, the relation
\begin{equation}
\zm_{5,7} = \frac{14}{9} \zm_{3,9} + \frac{28}{3} \zm_{5} \zm_{7} - \frac{ 121285 }{12438}\zm_{12}
\label{mzv12}
\end{equation}
is inaccessible from double-shuffle relations of depth two and requires
higher-depth input \cite{GKZ}.


\paragraph{Indecomposable eMZVs.} By applying the shuffle-reduction \eqn{mine3}
to higher-weight instances of the length-three Fay relation \eqn{sumexp2}, any
length-two eMZV can be expressed in terms of products of $\zm_{2k}$ and
$\omm(0,2n-1)$:
\begin{align}
  \omm(n_1,n_2) \, \Big|_{n_1+n_2 \ \te{odd}} &= (-1)^{n_1} \omm(0,n_1+n_2) + 2 \delta_{n_1,1} \zm_{n_2} \omm(0,1) - 2 \delta_{n_2,1} \zm_{n_1} \omm(0,1)\notag  \\
 & \ \ \ +2 \sum_{p=1}^{\lceil \frac{1}{2}(n_2-3) \rceil} {n_1+n_2 - 2p-2\choose n_1-1} \zm_{n_1+n_2-2p-1} \omm(0,2p+1) \label{mine7} \\
     & \ \ \ -2\sum_{p=1}^{\lceil \frac{1}{2}(n_1-3) \rceil} {n_1+n_2 - 2p-2\choose n_2-1} \zm_{n_1+n_2-2p-1} \omm(0,2p+1)\,,
 \notag
  \end{align}
which implies that no eMZVs at length two other than $\omm(0,2n-1)$ are
indecomposable. This relation can be straightforwardly proven using the
techniques of \subsecref{ssec:constq}.

Accordingly, the richest source of relations between interesting eMZVs at
length three are the length-four Fay relations at even weight together with the
shuffle reduction \eqn{sh42} of the boring eMZVs therein. The indecomposable
eMZVs can be chosen to include $\omm(0,0,2n)$ by analogy with \eqn{mine7}, and
additional indecomposable eMZVs such as $\omm(0,3,5)$ occur at weight
$w_\omega\geq 8$, e.g.
\begin{align}
\omm(1, 1, 2) &= \frac{13}{12}\zm_4 - 
  \zm_2 \omm(0, 1)^2 + \omm(0, 
     1) \omm(0, 3) + 
  3\zm_2 \omm(0, 0, 2) - 
  \frac{1}{2} \omm(0, 0, 4) 
  \label{fayl4}
  \\
\omm(0, 6, 2)&= -\frac{21}{2}\zm_8 +  2 \omm(0, 3) \omm(0, 5) - 
  14\zm_6 \omm(0, 0, 2) - 
  6\zm_4 \omm(0, 0, 4) - 
  \frac{9}{2} \omm(0, 0, 8) - \frac{2}{5} \omm(0, 3, 5) \ .
  \notag
\end{align}
Similarly, the set of indecomposable length-three eMZVs at weights ten and
twelve can be chosen as $\{\omm(0,0,10),\omm(0,3,7)\}$ and
$\{\omm(0,0,12),\omm(0,3,9)\}$, respectively. The weight-twelve relation
\begin{align}
\omm (0,5,7)=& -140 \zm_{10} \omm (0,0,2)-14 \zm_{8} \omm (0,0,4)+\frac{28}{3} \omm(0,5) \omm (0,7)\nnl
             & -\frac{119}{6} \omm (0,0,12)+\frac{14}{9} \omm (0,3,9)-\frac{550396}{6219} \zm_{12}\label{eqn:weight12examples}
\end{align}
will play an essential r\^ole later on. 

While even-weight single MZVs are special cases of length-one eMZVs by
\eqn{eqn:ellone}, odd MZVs do not show up in any relation for an eMZV of
$\ell_\om \leq 3$. When applying the above procedure to higher lengths,
$\zm_{3}$ is identified to be an eMZV by length-four relations such as
\begin{equation}
  \omm(0,1,2,0)= \frac{1}{4}\omm(0,3)-\frac{5}{2} \om (0,0,0,3)-\frac{\zm_{3}}{4} \ .
  \label{eqn:zeta3example}
\end{equation}
The appearance of $\zm_3$ in eMZV relations at length $\ell_\om =4,5$ is
governed by \eqn{lhs} at $r=4,5$, and similar relations are
expected to hold for any odd single zeta value by \eqn{extra21} and \eqn{tzero81}. Further
support stems from the description of the eMZVs' constant terms through the
Drinfeld associator~\cite{Drinfeld:1989st, Drinfeld2,Le} in \eqn{eqn:genconst}
below.

Usual MZVs show up in many relations between eMZVs such as
\eqn{eqn:zeta3example}. While crucial for matching the constant term for the
eMZVs in question, we will not count them as indecomposable eMZVs. Instead,
they will arise as suitably chosen boundary conditions for a differential
equation to be elaborated upon below. 

Table \ref{tab:emzvbasis} shows a possible (non-canonical) choice of
indecomposable eMZVs for weights up to 14 and length up to and including five.
The need for higher-length Fay relations increases the computational complexity
in the classification of indecomposable eMZVs using the above procedure. Hence,
comparing the $\tau$-dependence will enter as an additional method in the next
subsection to extend the results in the table to higher lengths and weights.
Still, shuffle, reflection and Fay relations were assembled completely at
$\ell_\om=2$, at $\ell_\om=3$ with $w_\om \leq 14$, at $\ell_\om=4$ with $w_\om
\leq 9$ as well as at $\ell_\om=5$ with $w_\om \leq 6$, and additional eMZV
relations at those weights and lengths have been ruled out on the basis of
their $q$-expansion.
\begin{table}[htp]
\setlength\tabcolsep{4pt}
\begin{minipage}[t]{0.45\textwidth}
\begin{tabular}{|c|| c|c|c|}
  \hline  \diagbox{$w_\om$}{$\ell_\om$} &2 &3 &4 \\ \hline \hline   
 1 &$\omm(0,1)$&&$\omm(0,0,1,0)$\\\hline
 3 &$\omm(0,3)$&&$\omm(0,0,0,3)$\\\hline
 5 &$\omm(0,5)$&&$\omm(0,0,0,5)$\\ 
   &&&$\omm(0,0,2,3)$\\\hline
 7 &$\omm(0,7)$&&$\omm(0,0,0,7)$\\ 
   &&&$\omm(0,0,2,5)$\\ 
   &&&$\omm(0,0,4,3)$\\\hline
 9 &$\omm(0,9)$&&$\omm(0,0,0,9)$\\ 
   &&&$\omm(0,0,2,7)$\\ 
   &&&$\omm(0,0,4,5)$\\ 
   &&&$\omm(0,1,3,5)$\\\hline
 11 &$\omm(0,11)$&&$\omm(0,0,0,11)$\\ 
    &&&$\omm(0,0,2,9)$\\ 
    &&&$\omm(0,0,4,7)$\\
    &&&$\omm(0,1,3,7)$\\
    &&&$\omm(0,3,3,5)$\\\hline
13  &$\omm(0,13)$&&$\omm(0,0,0,13)$\\ 
    &&&$\omm(0,0,2,11)$\\ 
    &&&$\omm(0,0,4,9)$\\ 
    &&&$\omm(0,1,3,9)$\\
    &&&$\omm(0,1,5,7)$\\
    &&&$\omm(0,3,3,7)$\\
    &&&$\omm(0,3,5,5)$\\\hline
\end{tabular}
\end{minipage}
\begin{minipage}[t]{0.55\textwidth}
\setlength\tabcolsep{4pt}
\begin{center}
\begin{tabular}{|c||  c|c|c|c|}
\hline \diagbox{$w_\om$}{$\ell_\om$} &2 &3 &4 &5  \\ \hline \hline   
 2 &&$\omm(0,0,2)$&&$\omm(0,0,0,0,2)$\\\hline
 4 &&$\omm(0,0,4)$&&$\omm(0,0,0,0,4)$\\ 
   &&&&$\omm(0,0,0,1,3)$\\\hline 
 6 &&$\omm(0,0,6)$&&$\omm(0,0,0,0,6)$\\ 
   &&&&$\omm(0,0,0,1,5)$\\
   &&&&$\omm(0,0,0,2,4)$\\
   &&&&$\omm(0,0,2,2,2)$\\\hline
 8 &&$\omm(0,0,8)$&&$\omm(0,0,0,0,8)$\\ 
   &&$\omm(0,3,5)$&&$\omm(0,0,0,1,7)$\\
   &&&&$\omm(0,0,0,2,6)$\\
   &&&&$\omm(0,0,1,2,5)$\\
   &&&&$\omm(0,0,2,2,4)$\\\hline
 10 &&$\omm(0,0,10)$&&$\omm(0,0,0,0,10)$\\ 
    &&$\omm(0,3,7)$&&$\omm(0,0,0,1,9)$\\
    &&&&and 7 more\\\hline
 12 &&$\omm(0,0,12)$&&$\omm(0,0,0,0,12)$\\
    &&$\omm(0,3,9)$&&$\omm(0,0,0,1,11)$\\
    &&&&$\omm(0,0,0,2,12)$\\
    &&&&and 11 more\\\hline
 14 &&$\omm(0,0,14)$&&$\omm(0,0,0,0,14)$\\
    &&$\omm(0,3,11)$&&$\omm(0,0,0,1,13)$\\
    &&$\omm(0,5,9)$&&$\omm(0,0,0,2,12)$\\
    &&&&and many more\\\hline
\end{tabular}
\end{center}
\end{minipage}
\caption{A possible choice of indecomposable eMZVs up to weight 14 and length
5. A table containing the elements missing here is available at \EMZVDatamine.}
\label{tab:emzvbasis}
\end{table}
Continuing the search for indecomposable eMZVs as described in previous and
subsequent subsections leads to \tabref{tab:basisdim}, in which the number of
indecomposable eMZVs for a certain length and weight are noted.
\begin{table}[htp]
\begin{center}
\begin{tabular}{|c||  c|c|c|c|c|  c|c|c|c|c|  c|c|c|c|c|  c|c|c|c|}
  \hline  \diagbox{$\ell_\om$}{$w_\om$} &1 &2 &3 &4 &5 &6 &7 &8 &9 &10 &11 &12 &13 &14 &15 &16 &17 &18 &19 \\ \hline \hline   
   2
&1 & &1 & &1
& &1 & &1 &
&1 & &1 & &1
& &1 &  &1   \\\hline
3  
& &1 & &1 &
&1 & &2 & &2
& &2 & &3 &
&3 &  &3 &  \\\hline
4
&1 & &1 & &2
& &3 & &4 &
&5 & &7 & &8
& &10  & &x  \\\hline
5
& &1 & &2 &
&4 & &6 & &9
& &13 & &x &
&x & &x  &  \\\hline
6  
&1 & &2 & &4
& &8 & &13 &
&x & &x & &x
& &x  & &x  \\\hline
7 
& &1 & &2 &
&x & &x & &x
& &x & &x &
&x &  &x &  \\\hline
%
\end{tabular}
\caption{Number $N(\ell_\om,w_\om)$ of indecomposable eMZVs at length
$\ell_\om$ and weight $w_\om$.}
\label{tab:basisdim}
\end{center}
\end{table}
Basis rules for rewriting each eMZV in terms of those indecomposable elements
can be obtained in digital form from the web page \EMZVDatamine{} and are
available up to and including weights $30,18,12,10$ for lengths $3,4,5,6$,
respectively. 


\subsection[Constant term and $q$-expansion]{Constant term and
\texorpdfstring{$\boldsymbol{q}$}{q}-expansion} \label{ssec:constq}

The system of relations discussed in the previous section did not require any
information on the eMZVs' functional dependence on the modular parameter
$\tau$. In this section, we determine the Fourier expansion in $q=e^{2\pi
i\tau}$ based on a first-order differential equation in $\tau$ along with a
boundary value at $\tau \rightarrow i\infty$. This will not only provide
crosschecks for the above eMZV-relations but will also lead to the more
efficient approach to classifying indecomposable eMZVs at higher length and
weight in \secref{sec:gamma}.


\paragraph{Constant term.} The constant term of an eMZV can be determined
explicitly using results of ref.~\cite{EnriquezEMZV}. By construction, the
elliptic KZB associator $A(\tau)$ is the generating series of eMZVs, 
\begin{equation}
e^{\pi i[y,x]}A(\tau)\equiv \sum_{r \geq 0} (-1)^r \sum_{n_1,n_2,\ldots,n_r\geq 0} \omega(n_1,n_2,\ldots,n_r)\ad^{n_r}_x(y)\ldots \ad^{n_2}_x(y) \ad^{n_1}_x(y) \ ,
\label{nils1}
\end{equation}
and captures the monodromy of the elliptic KZB equation \cite{KZB, Hain} along
the path $[0,1]$. The prefactor $e^{\pi i[y,x]}$ is adjusted to the
regularization scheme in \eqn{eqn:shufreg}. The variables $x$ and $y$ generate
a complete, free algebra $\ZC \langle \!\langle x,y \rangle \! \rangle$ of
formal power series with complex coefficients, whose multiplication is the
concatenation product, and the convention for the adjoint action is 
\begin{equation}
 \ad_x(y) \equiv [x,y] \co \ad_x^n(y) = 
\underbrace{[ x, \ldots[x, [x,}_{n\ \te{times}}
y]]\ldots ]
 \ .
 \label{nils0}
\end{equation}
Note that the appearance of eMZVs in \eqn{nils1} along with non-commutative
words in $x$ and $y$ allows for an alternative enumeration scheme using a
two-letter alphabet, see \subsecref{ssec:prerequisites}.

Enriquez proved that $A(\tau)$ admits the asymptotic expansion as $\tau \to
i\infty$ \cite{EnriquezEMZV}
\begin{equation}
  A(\tau)=\Phi(\tilde{y},t)\,e^{2\pi i\tilde{y}}\,\Phi(\tilde{y},t)^{-1}+{\cal{O}}(e^{2\pi i\tau}) \ ,
  \label{AAssociator}
\end{equation}
where ${\cal{O}}(e^{2\pi i\tau})$ refers to the non-constant terms in
\eqn{eqn:qexp} exclusively. In the above equation, the genus-one alphabet
consisting of $x,y$ is translated into a genus-zero alphabet involving  
\begin{equation}
t\equiv [y,x] \co \tilde{y}\equiv -\frac{\ad_x}{e^{2\pi i\ad_x}-1}(y) \ ,
\label{alph}
\end{equation}
and $\Phi$ denotes the Drinfeld associator \cite{Drinfeld:1989st, Drinfeld2,
Le} 
\begin{equation}
\Phi(e_0,e_1) \equiv \sum_{\hat W \in \langle e_0,e_1 \rangle} \zeta^{\shuffle}(W) \cdot \hat W \ .
\end{equation}
The sum over $\hat W \in \langle e_0,e_1 \rangle$ includes all non-commutative
words in letters $e_0$ and $e_1$, and the word $W$ is obtained from $\hat W$ by
replacing letters $e_0$ and $e_1$ by $0$ and $1$, respectively. Then,
$\zeta^{\shuffle}(W)$ denote shuffle-regularized MZVs \cite{Rac} which are
uniquely determined from \eqn{mzvdef}, the shuffle product
\eqn{eqn:zetashuffle} and the definition
$\zeta^{\shuffle}(0)=\zeta^{\shuffle}(1)=0$ for words of length one.
Consequently, the first few terms of $\Phi(e_0,e_1)$ are given by 
\begin{equation}
\Phi(e_0,e_1)=1-\zm_2 [e_0,e_1]-\zm_3 [e_0+e_1,[e_0,e_1]] + \ldots \ .
\end{equation}
From \eqns{nils1}{AAssociator}, the generating
series of constant terms $\omm_0(n_1,...,n_r)$ of eMZVs is immediately obtained as
\begin{equation}
\sum_{r \geq 0} (-1)^r \sum_{n_1,\ldots,n_r\geq 0} \omm_0(n_1,\ldots,n_r)\ad^{n_r}_x(y)\ldots\ad^{n_1}_x(y)=e^{\pi i [y,x]}\,\Phi(\tilde{y},t)\,e^{2\pi i\tilde{y}}\,\Phi(\tilde{y},t)^{-1} \ .
\label{eqn:genconst}
\end{equation}
In order to transfer information from the right hand side of \eqn{eqn:genconst}
to the constant terms of eMZVs on the left hand side, it remains to expand
words in the alphabet $\{\tilde{y},t\}$ in \eqn{alph} as formal series of words
in the alphabet $\{\ad^n_x(y) \, \vert \, n \geq 0\}$ and then to compare the
coefficients of both sides.

Perhaps surprisingly, the case where all $n_i \neq 1$ is very simple to treat.
In that case, only the middle term $e^{2\pi i\tilde{y}}$ from
\eqn{eqn:genconst} yields a non-trivial contribution, and therefore we have
\begin{align}
\omega_0(n_1,n_2,\ldots,n_r) \Big|_{n_i \neq 1}&=\left\{
\begin{array}{cl}
0 & \mbox{if at least one $n_i$ is odd, and all $n_i \neq 1$}\\
\displaystyle \frac{1}{r!}\prod_{i=1}^r(-2\zm_{n_i}) & \mbox{if all $n_i$ are even}
\end{array}
\right. \,.
\label{eqn:constnoones}
\end{align}
In particular, one finds 
\begin{equation}
  \omm(\underbrace{0,0,\ldots,0}_{n\ \te{times}})=\frac{1}{n!} \ ,
  \label{eqn:omzero}
\end{equation}
which is perfectly in line with $f^{(0)} \equiv 1$.
On the other hand, in presence of $n_i=1$ at some places, a general formula for the 
constant term is very cumbersome. Simple instances include
\begin{align}
&\omm_0(1,0) = -\frac{i\pi}{2} \co \omm_0(1,0,0) = -\frac{i\pi}{4}  \co \omm_0(1,0,0,0) = -\frac{i \pi }{12} - \frac{\zm_{3}}{24 \zm_{2}}
 \notag \\
&\ \ \ \ \omm_0(0, 1, 1, 0, 0) =\frac{ \zm_{2}}{15} \co
\omm_0(1, 0, 1, 1, 0, 0) =
-\frac{i \pi \zm_2}{30}  -\frac{\zm_3}{8} - \frac{17  \zm_{5}}{96 \zm_2} 
\end{align}
with generalizations in \eqn{constw1}. Replacing $n_i=0$ in the
above identities by even values $n_i=2k$ amounts to multiplication with
$-2\zm_{2k}$ on the right hand side.


\paragraph{\texorpdfstring{$\boldsymbol{q}$}{q}-expansion.} The $q$-dependent
terms in the expansion can be determined using the known form of the
$\tau$-derivative of eMZVs. In Th\'eor\`eme 3.3 of ref.~\cite{EnriquezEMZV},
the derivative of a generating functional for eMZVs is presented, which
translates as follows into derivatives of individual eMZVs in our conventions:
\begin{align}
2\pi i &\frac{\dd}{\dd \tau} \omm(n_1,\ldots,n_r) =-4\pi^2 q \frac{\dd}{\dd q} \omm(n_1,\ldots,n_r) \notag \\
&=  n_1 \GG{n_1+1} \omm(n_2,\ldots,n_r) - n_r \GG{n_r+1} \omm(n_1,\ldots,n_{r-1})\notag \\
&+ \sum_{i=2}^r \Big\{ (-1)^{n_i} (n_{i-1}+n_i)   \GG{n_{i-1}+n_i+1} \omm(n_1,\ldots,n_{i-2},0,n_{i+1},\ldots,n_r) \label{eqn:tauder} \\
& \ \ \ \ \ \ -  \sum_{k=0}^{n_{i-1}+1} (n_{i-1}-k) { n_i+k-1 \choose k }  \GG{n_{i-1}-k+1} \omm(n_1,\ldots,n_{i-2},k+n_i,n_{i+1},\ldots,n_r) \notag \\
& \ \ \ \ \ \ + \sum_{k=0}^{n_{i}+1} (n_{i}-k) { n_{i-1}+k-1 \choose k }  \GG{n_{i}-k+1} \omm(n_1,\ldots,n_{i-2},k+n_{i-1},n_{i+1},\ldots,n_r) \Big\} \ .
\notag 
\end{align}
The \emph{Eisenstein series} $\GG{k}\equiv\GG{k}(\tau)$ on the right hand side
are defined by\footnote{The case $k=2$ requires the Eisenstein summation
prescription
\[
	\sum_{m,n \in \mathbb Z}a_{m,n}\equiv
	\lim_{N \to \infty}\lim_{M \to \infty} \sum_{n=-N}^N\sum_{m=-M}^Ma_{m,n} \ .
\]
}
\begin{align}
  \GG{0}(\tau)&\equiv-1
\notag \\
\GG{k}(\tau)&\equiv \left\{ \begin{array}{cl} \displaystyle \sum\limits_{m,n \in \mathbb Z \atop{(m,n) \neq (0,0)}} \frac{1}{(m+n\tau)^k} &: \ k>0\ \te{even} \ ,\\
0&: \ k>0\ \te{odd}\ . \end{array} \right. 
	     \label{eqn:eiss}
\end{align}
Positive even values of $k$ admit a series expansion in the modular parameter:
\begin{equation}
  \GG{k}(\tau)= 2\zm_k+\frac{2(-1)^{k/2}(2\pi)^k}{(k-1)!} \sum_{m,n=1}^{\infty} m^{k-1} q^{mn} \qquad k>0\ \te{even}\,.
  \label{eqn:eisexp}
\end{equation}
Using the above formul\ae{} and the known expansion of the Eisenstein series
$\GG{k}$ in \eqn{eqn:eisexp}, one can recursively obtain the explicit
$q$-expansion for any eMZV: The length of eMZVs on the right hand side of
\eqn{eqn:tauder} is decreased by one compared to the left hand side, and the
recursion terminates with the constant eMZVs at length one given by
\eqn{eqn:ellone}. 

In addition, one finds from \eqn{eqn:tauder} that only divergent eMZVs with
$n_1=1$ or $n_r=1$ lead to the non-modular $\GG{2}$, see the discussion around
\eqn{eqn:shufreg}. In all other situations which lead to the non-modular
$\GG{2}$ in the last three lines, the respective terms cancel out.  Not
surprisingly, the interesting and boring character of eMZVs is preserved by
\eqn{eqn:tauder}: the decreased length on the right hand side is compensated by
an increased weight.

Also, note that the differential equation \eqn{eqn:tauder} contains no MZV
terms. In fact, the only way through which MZVs enter the stage of eMZVs is by
means of the constant term \eqn{AAssociator} of the KZB associator. As
mentioned earlier, this constant term can be thought of as a boundary-value
prescription for the differential equation \eqn{eqn:tauder}, thereby
determining eMZVs uniquely.


\paragraph{eMZV relations from the $q$-expansion.} Based on the $q$-expansions
described above, relations between eMZVs can be checked and ruled out by comparing their Fourier representations.
In practice, one writes down an ansatz comprised from interesting eMZVs and
products thereof with uniform weight and an upper bound on the length of
interest, each term supplemented with fudge coefficients. 

Na\"ively, one could calculate the $q$-expansions of all constituents up to a
certain order $q^{N_{\te{max}}}$ and impose a matching along with each Fourier
mode $q^n$ for $0 \leq n \leq N_{\te{max}}$. This allows to fix the above fudge
coefficients and to check the relations' validity up to -- in principle --
arbitrary order. In an early stage of the project, our computer implementation
of this approach with $N_{\te{max}}=160$ was far more efficient compared to the
analysis of reflection, shuffle and Fay identities and lead to substantial
parts of the data shown in tables \ref{tab:emzvbasis} and \ref{tab:basisdim}. 

However, since the comparison of $q$-expansions has to be cut off at some
chosen power of $q$, a proof for relations using the method is impossible by
construction. Even worse, this na\"ive method fails to capture the structural
insight from \eqn{eqn:tauder} that any $q$-dependence in eMZVs stems from
iterated integrals of Eisenstein series. This crucial property is exploited in
\secref{sec:gamma}, confirming the entries of \tabref{tab:basisdim} in a
rigorous and conceptually by far more elegant manner.

While the na\"ive comparison of Fourier coefficients merely provides a lower
bound for the number of indecomposable eMZVs for a given weight and length, the
description of eMZVs in terms of iterated Eisenstein integrals in
\secref{sec:gamma} yields complementary upper bounds.  Under the additional
assumption that different iterated Eisenstein integrals are linearly
independent, these upper bounds are indeed saturated. However, since we do not
attempt to prove their linear independence, the na\"ive matching of
Fourier coefficients closes the associated loophole at the weights and lengths
under consideration.


\section{Multiple zeta values and the \texorpdfstring{$\boldsymbol{\phi}$}{p}-map}
\label{sec:MZV}

In this section we gather information on the structure of MZVs, which are to be
compared with those found for eMZVs in \secref{sec:gamma} below. While
represented as nested sums in \eqn{mzvsum} in \secref{sec:eMZVrel}, they can
alternatively be defined as iterated integrals
\begin{align}
\zm_{n_1,n_2,\ldots,n_r} 
&= \int\limits_{0\leq z_i\leq z_{i+1}\leq 1} \om_1 \underbrace{\om_0 \ldots \om_0}_{n_1-1} \om_1 \underbrace{\om_0 \ldots \om_0}_{n_2-1} \ldots \om_1 \underbrace{\om_0 \ldots \om_0}_{n_r-1}\nnl
&=\zm(1 \underbrace{0 \ldots 0}_{n_1-1} 1 \underbrace{0 \ldots 0}_{n_2-1} \ldots 1 \underbrace{0 \ldots 0}_{n_r-1})
\label{mzvdef}
\end{align}
over the differential forms $\om_0 \equiv \frac{\dd z}{z}$ and $\om_1 \equiv
\frac{\dd z}{1-z}$ with all $z_i$ on the real line. The MZV
$\zeta_{n_1,...,n_r}$ is said to have \textit{weight} $w=\sum_{i=1}^r n_i$ and
\textit{depth} $r$. Written in terms of words $W$ composed from the letters 0
and 1, which correspond to the differential forms $\om_0$ and $\om_1$ in
\eqn{mzvdef}, respectively, $\zm$'s satisfy the shuffle product:
\begin{equation}
  \zm(W_1)\zm(W_2)=\zm(W_1\shuffle W_2)\,.
  \label{eqn:zetashuffle}
\end{equation}
There is also a second product structure on MZVs, the stuffle
product. Its simplest instance reads
\begin{equation}
  \zm_m\,\zm_n = \zm_{m,n}+\zm_{n,m}+\zm_{m+n}.
  \label{qshuffle}
\end{equation}
It follows from either \eqn{eqn:zetashuffle} or \eqn{qshuffle} that the
$\ZQ$-span $\CZ$ of all MZVs is a subalgebra of $\ZR$. Conjecturally, $\CZ$ is graded by the weight of the MZVs
\begin{equation}
  \CZ=\bigoplus_{w=0}^{\infty} \CZ_w\,,
  \label{Zalgebra}
\end{equation}
where the dimensions $d_w$ of $\CZ_w$ have been conjectured to be
$d_w=d_{w-2}+d_{w-3}$ where $d_0=1$, $d_1=0$ and $d_2=1$ \cite{Zagier23}.
A possible choice of basis elements for each weight $w$ is given
in \tabref{zetaBasis}, for higher weights consult ref.~\cite{Blumlein:2009cf}.

\begin{table}[h]
  \[
\begin{array}{|l|l|l|l|l|l|l|l|l|l|ll|ll|}
\noalign{\hrule}
 w &2 &3 & 4  & 5 & 6& 7 & 8 & 9 & 10 & 11 &&12&\nnl
\noalign{\hrule}
\CZ_w & \zeta_2  &\zeta_3  & \zeta_2^2 &\zeta_5 & \zeta_3^2 &\zeta_7  &\zeta_{3,5} &\zeta_9 &\zeta_{3,7} & \zeta_{3,3,5}&\zeta_2\  \zeta_3^3 & \zeta_{1,1,4,6}&\zeta_2\ \zeta_{3,7} \nnl
\ &\  &\  & \  &\zeta_2\ \zeta_3 & \zeta_2 ^3 &\zeta_2\ \zeta_5  &\zeta_3\ \zeta_5 &\zeta_3^3&\zeta_3\ \zeta_7 &\zeta_{3,5}\ \zeta_3&\zeta_2\ \zeta_9  
&\zeta_{3,9}&\zeta_2^2\ \zeta_{3,5}\nnl
\ &\  &\  & \  &\ &\ & \zeta_2^2\ \zeta_3  &\zeta_2\ \zeta_3 ^2&\zeta_2\ \zeta_7 &\zeta_5^2&\zeta_{11}& \zeta_2^2\ \zeta_7  &\zeta_3\ \zeta_9&
\zeta_2\ \zeta_5^2\nnl
\ &\  &\  & \  &\ &\ &\  &\zeta_2^4&\zeta_2^2\ \zeta_5 &\zeta_2\ \zeta_{3,5} &\zeta_3^2\ \zeta_5&\zeta_2^3\ \zeta_5   &\zeta_5\ \zeta_7&\zeta_2\ \zeta_3\ \zeta_7\nnl
\ &\  &\  & \  &\ &\ &\  &\ &\zeta_2^3\ \zeta_3 &\zeta_2\ \zeta_3\ \zeta_5  &   &\zeta_2^4\ \zeta_3   &\zeta_3^4&\zeta_2^2\ \zeta_3\ \zeta_5\nnl
\ &\  &\  & \  &\ &\ &\  &\ &\ &\zeta_2^2\ \zeta_3^2 & &  & &\zeta_2^3\ \zeta_3^2 \nnl
\ &\  &\  & \  &\ &\ &\  &\ &\ &\zeta_2^5 & &  & & \zeta_2^6 \nnl
\noalign{\hrule}
d_w &1 &1  &1 &2 &2 &3  &4 &5 &7 & 9& &12& \nnl
\noalign{\hrule}
\end{array}
\]
\caption{A possible choice for the basis elements of $\CZ_w$ for $2\leq w\leq 12$.}
\label{zetaBasis}
\end{table}

Single $\zm$-functions of even weight are rather different from their
odd-weight counterparts: all single zeta values of even weight $2n$ can be
expressed as rational multiples of $\pi^{2n}$, which renders them
transcendental numbers immediately. For odd single zeta values, however, there
is no analogous property: there are no known relations relating two single zeta
values of distinct odd weight, and in fact no such relations are expected.
Also, although expected, none of the odd $\zm$-values has been proven to be
transcendental so far: the only known facts are the irrationality of $\zm_3$ as
well as the existence of an infinite number of odd irrational $\z$'s
\cite{Apery,BallRivoal}.


\subsection{Hopf algebra structure of MZVs} 

The basis elements in \tabref{zetaBasis} have been chosen by convenience
preferring short and simple $\zm$'s. However, the choice of basis elements does
not seem to be intuitive at all, as is exemplified by the appearance of
$\zm_{1,1,4,6}$ at weight $12$. It would be desirable to find a language in
which one can write down a basis for MZVs in a more transparent way, with all
relations built in once the translation is performed. This language does indeed
exist: it is furnished by the graded Hopf algebra comodule $\CU$, which is
composed from words
\begin{equation}
  f_{2i_1+1}\ldots f_{2i_r+1}\ f_2^k, \quad\text{with}\quad r,k\geq 0 \quad\text{and}\quad i_1,\ldots,i_r\geq 1
  \label{UBasis}
\end{equation}
of weight $w=2 (i_1+\ldots+i_r)+r+2k$. While words in the letters $f_{2i+1}$
span a Hopf algebra endowed with a commutative shuffle product, the Hopf
algebra comodule $\CU$ is obtained upon adjoining powers of $f_2$, which
commute with all $f_{2i+1}$ \cite{BrownTate}.  Writing down all words of the
form in \eqn{UBasis}, one indeed finds the dimension of $\CU_w$ to match the
expected dimension $d_w$ of $\CZ_w$, which is a first indicator that the Hopf
algebra comodule $\CU$ does indeed shed light on the algebraic structure of
MZVs.

In a next step MZVs need to be related to elements in $\CU$. Unfortunately, due
to the difficult problem of excluding algebraic relations between MZVs,
this cannot be done directly. In order to circumvent this issue, one lifts MZVs
$\zm$ to so-called motivic MZVs $\zm^\fm$, which have a more elaborate
definition \cite{Goncharov:2005sla, BrownTate, Brown:ICM14}, but which still
satisfy the same shuffle and stuffle product formul\ae{} as the MZVs
\eqns{eqn:zetashuffle}{qshuffle}.  Moreover, passing from MZVs to motivic MZVs
has the advantage that many of the desirable, but currently unproven facts
about MZVs are in fact proven for motivic MZVs.  In particular, the commutative
algebra $\CH$ of motivic multiple zeta values is by definition graded for the
weight, and carries a well-defined motivic coaction, first written down by
Goncharov \cite{Goncharov:2005sla} and further studied by Brown
\cite{BrownTate,Brown:2011ik,Brown:ICM14}.

With the availability of $\CH$ the only remaining piece is the construction of
an isomorphism $\phi$ of graded algebra comodules
\begin{equation}
  \phi:\CH\rightarrow\CU\,,
  \label{Phimap}
\end{equation}
whose existence is guaranteed by the main result of \cite{BrownTate}.  The map
$\phi$, which assigns to each motivic MZV a linear combination of the words
defined in \eqn{UBasis}, is thoroughly described and explored in
ref.~\cite{Brown:2011ik}. As pointed out in the reference, the map $\phi$ is
non-canonical and depends on the choice of an algebra basis of $\CH$. The
requirement that all odd single motivic MZVs as well as $\zm_2$ should be
contained in this basis leads to
\begin{equation}
 \phi(\zm_k^\fm)=f_k\,,\quad k=2,3,5,7,\ldots\,.
\end{equation} 
Unfortunately, this convention does not determine $\phi$ uniquely, since not
every motivic MZV can be expressed in terms of motivic single zeta values only.
However, as pointed out in ref.~\cite{Brown:2011ik}, the map $\phi$ preserves
all relations between motivic MZVs for any choice of algebra basis of $\CH$,
for example~(cf. \eqn{qshuffle}):
\begin{equation}
  \phi(\z^\fm_m\,\z^\fm_n)=\phi(\z^\fm_{m,n})+\phi(\z^\fm_{n,m})+\phi(\z^\fm_{m+n})\,.
  \label{nettesBeispiel}
\end{equation}
In order to give explicit examples of the decomposition of motivic MZVs into
the $f$-alphabet, let us choose the following algebra basis up to weight $12$,
upon which \tabref{zetaBasis} is modeled implicitly
\begin{equation}
\{\zm_2^\fm,\,\zm_3^\fm,\,\zm_5^\fm,\,\zm_7^\fm,\,\zm_{3,5}^\fm,\,\zm_9^\fm,\,\zm_{3,7}^\fm,\,\zm_{11}^\fm,\,\zm_{3,3,5}^\fm,\,\zm_{1,1,4,6}^\fm,\,\zm_{3,9}^\fm\}\,.
\end{equation}
With this choice of basis, one finds for example,
\begin{align}
  \phi(\z^\fm_{3,9}) &= -6\,f_5 f_7 - 15\,f_7 f_5 - 27\,f_9 f_3 \nnl
  \phi(\z^\fm_{3,3,5}) &=-5\,f_5f_3f_3 + \frac{4}{7}\,f_5f_2^3-\frac{6}{5}\,f_7f_2^2-45\,f_9f_2\,.
  \label{phiExamples}
\end{align}
The application of the $\phi$-map in the context of the low-energy expansion of
superstring tree-level amplitudes as well as several higher-weight examples can
be found in ref.~\cite{Schlotterer:2012ny}.


\section{Indecomposable eMZVs, Eisenstein series and the derivation algebra}
\label{sec:gamma}

As described in \secref{sec:eMZVrel}, indecomposable eMZVs at a certain weight
and length can be in principle inferred from considering reflection, shuffle
and Fay relations. For higher weights and lengths,
however, it is favorable to employ a computer implementation based on
comparing $q$-expansions of eMZVs which in turn can be obtained recursively
from \eqn{eqn:tauder}. In this section we are going to provide an algorithm
which does not only deliver the appropriate indecomposable elements as listed
in table \ref{tab:emzvbasis} but as well explains their number at a given
length and weight. 

As described in the previous section, the appropriate mathematical idea for
standard motivic MZVs is to map them to the non-commutative words composed from
letters $f_w$ in \eqn{UBasis} using the map $\phi$. For the elliptic case we
will construct an isomorphism $\psi$ relating the
$\om$-representation of eMZVs to non-commutative words composed from letters
$g_w$, which in turn arise as labels of iterated Eisenstein integrals $\gm$ to
be defined below. 


\subsection{Iterated Eisenstein integrals}
\label{ssec:fourzero}

Given that the $q$-expansion of eMZVs can be iteratively generated from the
Eisenstein series $\GG{k}$ employing \eqn{eqn:tauder}, we will now describe
eMZVs based on combinations of $\GG{k}$.  Instead of representing eMZVs as
elliptic iterated integrals as in \secref{sec:eMZVrel}, we will write them as
iterated integrals over Eisenstein series $\GG{k}$ where the iterated
integration is now performed over the modular parameter $\tau$ (or equivalently
$q$). 

Iterated integrals over Eisenstein series arise as a subclass of iterated
integrals of modular forms, which have been studied in
refs.~\cite{ManinIterMod,Brown:mmv}.  In this section, we will briefly review
some of the key definitions in order to embed the subsequent presentation of
eMZVs into a broader context.

Iterated integrals of modular forms or {\em iterated Shimura integrals}
\cite{ManinIterMod,Brown:mmv} are defined via
\begin{align}
 \int\limits_{i \infty > \tau_1 >
\tau_2 > \cdots > \tau} \!\!\!\!\dd \tau_1\ (X_1-\tau_1 Y_1)^{k_1-2} {\cal F}_{k_1}(\tau_1) \ 
            \dd \tau_2 & \ (X_2-\tau_2 Y_2)^{k_2-2} {\cal F}_{k_2}(\tau_2)\cdots \nnl
	    &\cdots \dd \tau_n\  (X_n-\tau_n Y_n)^{k_n-2}{\cal F}_{k_n}(\tau_n)  \ ,
\label{modint} 
\end{align}
where ${\cal F}_{k}(\tau)$ is a modular form of weight $k$ and the modular
group acts on commutative variables $X_i$ and $Y_i$ as to render \eqn{modint}
modular invariant. The divergences in these integrals caused by the constant
terms in the $q$-expansion of the modular forms can be regularized in a manner
described in ref.~\cite{Brown:mmv}. The key idea of this regularization procedure is
to separate the constant part from the remaining $q$-series for each ${\cal
F}_{k_j}(q)$ and to associate a different integration prescription to it. The
mathematical justification of this procedure is furnished by the theory of
\textit{tangential base points} \cite{Deligne}. In the present case, one
regularizes the integral with respect to the tangential base point
$\vec{1}_{\infty}$ \cite{Brown:mmv}.

In the context of eMZVs in \eqn{l2.7}, we encounter special cases of the
iterated Shimura integrals defined above, evaluated at $X_i=1$ and $Y_i=0$.
Furthermore, the $\tau$-derivative of eMZVs in \eqn{eqn:tauder} involves no
modular forms ${\cal F}_{k}$ other than Eisenstein series $\GG{k}$. This
motivates to study the following {\em iterated Eisenstein integrals} as 
building blocks for eMZVs, 
\begin{align}
&\gm(k_1,k_2,\ldots,k_n;q) \equiv
\frac{1}{4\pi^2} \int_{0 \leq q'\leq q} \dlog q' \ \gm(k_1,\ldots,k_{n-1};q') \GG{k_n}(q') 
\label{eqn:defEInew} \nnl
&\ \ \ \ = \frac{1}{(4\pi^2)^n} \int_{0 \leq q_i<q_{i+1} \leq q} \dlog q_1\  \GG{k_1}(q_1) \ \dlog q_2 \ \GG{k_2}(q_2)\ \ldots \ \dlog q_n \GG{k_n}(q_n)  \ ,
\end{align}
where the number $n$ of integrations will be referred to as the \textit{length}
$\ell_\ga$, and the \textit{weight} is given by $w_\ga=\sum_{i=1}^n\,k_i$. The
definition in \eqn{eqn:defEInew} as an iterated integral immediately implies
\begin{align}
\frac{ \dd }{\dlog q} \gm(k_1,k_2,\ldots,k_n;q) &= \frac{ \GG{k_n}(q) }{4\pi^2} \gm(k_1,k_2,\ldots,k_{n-1};q)
\label{eqn:diffgam} \\
\gm(n_1,n_2,\ldots,n_r;q) \gm(k_1,k_2,\ldots,k_s;q) &= \gm \!  \big( (n_1,n_2,\ldots,n_r) \shuffle (k_1,k_2,\ldots,k_s) ;q
\big)\, ,
\label{gammash}
\end{align}
where the dependence on $q$ will be suppressed in most cases:
$\gm(\ldots)\equiv\gm(\ldots;q)$. The integrals in \eqn{eqn:defEInew} generally
diverge due to the constant term in $ \GG{k_1}= 2 \zm_{k_1} + {\cal O}(q)$ and
can be regularized using the procedure discussed around \eqn{REG} while
preserving \eqns{eqn:diffgam}{gammash}.

\begin{figure}
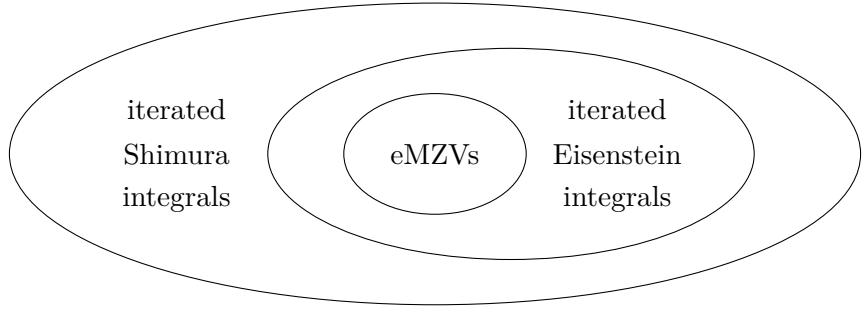

  \begin{center}
\tikzpicture[scale=0.4]
\draw (0,0)node{eMZVs};
\draw(6,1.5)node{iterated};
\draw(6,0)node{Eisenstein};
\draw(6,-1.5)node{integrals};
\draw(-8.5,1.5)node{iterated};
\draw(-8.5,0)node{Shimura};
\draw(-8.5,-1.5)node{integrals};
\draw (0,0) ellipse (3cm and 2cm);
\draw (2.5,0) ellipse (8cm and 3.5cm);
\draw (0,0) ellipse (14cm and 5cm);
\endtikzpicture
\end{center}
\caption{Relation between different type of iterated integrals discussed.}
\label{fig:integrals}
\end{figure}

As will be explained in detail below, eMZVs can be expressed in terms of
particular linear combinations of iterated Eisenstein integrals in
\eqn{eqn:defEInew} such that all possible divergences cancel. An alternative
description of eMZVs which manifests the absence of divergences and admits
convenient formul\ae{} for their $q$-expansion will be given in subsection
\ref{sec:simple}. The convergent linear combinations of \eqn{eqn:defEInew}
occurring in eMZVs will turn out to be governed by a special derivation algebra
$\DAlg$. The situation is summarized in \figref{fig:integrals}: eMZVs are a
special case of iterated Eisenstein integrals \eqn{eqn:defEInew} which in turn
span a subspace of iterated Shimura integrals \eqn{modint}.
%


\paragraph{Regularization.} Even though eMZVs can be assembled from convergent
iterated integrals over modular parameters -- see \subsecref{sec:simple} -- we
shall sketch a regularization procedure for the iterated Eisenstein integrals
in \eqn{eqn:defEInew} to render individual terms in the subsequent description
of eMZVs well-defined. Let us consider the simplest case, namely that of an
iterated integral of length one:
\begin{equation}
\gamma(k)=\frac{1}{4\pi^2}\int_{0}^{q}\GG{k}(q')\dlog q'.
\label{IterEis:lengthone}
\end{equation}
The term $(\GG{k}(q')-2\zm_k)\,\dlog q'$ is straightforward to integrate from
$0$ to $q$, since it has no poles on the integration domain $0 \leq q' \leq q$.
On the other hand, integration of the term $2\zm_k\dlog q'$ in isolation
requires regularization, due to the presence of a simple pole at $q'=0$. The
regularization scheme employed in this case, however, is entirely analogous to
the regularization scheme for multiple polylogarithms, MZVs or eMZVs: One
introduces a small parameter $\varepsilon >0$, then expands the integral
\begin{equation}
2\zm_k\int_{\varepsilon}^q \dlog q' = 2 \zm_k(\log q - \log \varepsilon)
\end{equation}
as a polynomial in $\log(\varepsilon)$ and finally takes the constant term in
this expansion. Using this procedure in the length one case, one obtains from
\eqn{IterEis:lengthone}
\begin{equation}
\gm(k)=\frac{1}{4\pi^2}\int_{0}^{q}\GG{k}(q')\dlog q'=\frac{1}{4\pi^2}\left(2\zm_k\log q+\frac{2(-1)^{k/2}(2\pi)^k}{(k-1)!} \sum_{m,n=1}^{\infty} \frac{m^{k-2}}{n} q^{mn}\right)\,.
\label{REG}
\end{equation}
The regularization procedure for a general iterated Eisenstein integral $\gamma(k_1,k_2,\ldots,k_n)$ as in \ref{eqn:defEInew} is deduced from the length one case, using the shuffle product formula. Full details can be found in \cite{Brown:mmv}.


\subsection{eMZVs as iterated Eisenstein integrals}
\label{ssec:eisint}

As a first example on how to express eMZVs in terms of iterated Eisenstein
integrals, let us consider \eqn{eqn:tauder} for two simple types of eMZVs
(recalling \eqn{eqn:qtau} and $\GG{0} \equiv -1$):
\begin{subequations}
\begin{alignat}{3}
2\pi i \frac{\dd}{\dd \tau}\omm(0,n) =-4\pi^2 q\frac{\dd}{\dd q}\omm(0,n) &= 
-2n \zm_{n+1}\GG{0}-n\GG{n+1},\quad &&n \ \te{odd}
\label{eqn:tau3} \\
2\pi i \frac{\dd}{\dd \tau}\omm(0,0,n) =-4\pi^2 q\frac{\dd}{\dd q}\omm(0,0,n) &=  n \omm(0,n+1) \GG{0} \ , &&n \ \te{even} \ .
\label{eqn:tau4}
\end{alignat}
\end{subequations}
Integration over $\dlog q$ relates the eMZVs on the left hand side to
iterated Eisenstein integrals of the form \eqn{eqn:defEInew}, and the absence
of constant terms within $\tau$-derivatives guarantees that the integral
converges. This insight will actually be the key ingredient to the simplified
representation of eMZVs described in \subsecref{sec:simple} below. The
rewriting in \eqns{eqn:tau3}{eqn:tau4} can be generalized for all eMZVs: using
the differential equation (\ref{eqn:tauder}) one can represent their derivative
as a sum over Eisenstein series $\GG{2k}$,
\begin{equation}
  \frac{\dd }{\dlog q} \omm(n_1,n_2,\ldots,n_r)= \frac{1}{4\pi^2} \sum_{k=0}^{\infty} \xi_{2k}(n_1,n_2,\ldots,n_r) \GG{2k}\, ,
\label{logq2}
\end{equation}
where the coefficients $\xi_{2k}(n_1,\ldots,n_{r})$ are linear combinations of
eMZVs of weight $n_1+\ldots+n_r+1-2k$ and length $r-1$. An example of this
decomposition is spelled out below \eqn{eqn:der1G}.

For the eMZVs appearing in the coefficients $\xi_{2k}(n_1,\ldots,n_{r})$ of
\eqn{logq2}, the procedure can be repeated to successively reduce the length.
Finally, any eMZV can be rewritten in terms of the \textit{iterated Eisenstein
integrals} in \eqn{eqn:defEInew}. Since the right hand side of \eqn{logq2} is 
a $\tau$-derivative and cannot have a constant term in $q$, its integral over $\dd
\log q$ is convergent and the first entries of the resulting iterated
Eisenstein integrals for any eMZV are interlocked as $\gm(k,\ldots)+2 \zeta_k
\gm(0,\ldots)$. 


\paragraph{Examples.} Let us return to the examples \eqns{eqn:tau3}{eqn:tau4}.
The differential equation \eqn{eqn:diffgam} immediately implies
\begin{subequations}
\begin{alignat}{3}
  \omm(0,n)&=\delta_{1,n}\frac{\pi i}{2}+n\big(\gm(n+1)  +2\zm_{n+1} \gm(0)\big),\quad&&n \ \te{odd}
\label{eqn:gamma1G} \\
\omm(0,0,n)&=-\frac{1}{3}\zm_n - n(n+1) \big(\gm(n+2,0)+2\zm_{n+2}\gm(0,0)\big),\quad&&n \ \te{even}\,,
\label{eqn:gamma2G}
\end{alignat}
\end{subequations}
where $\delta_{1,n}\frac{\pi i}{2}$ and $-\frac{1}{3}\zm_n$ arise as integration constants w.r.t.~$\log q$.
Even though all the above iterated Eisenstein integrals
$\gm(n+1),\gm(n+2,0),\gm(0)$ and $\gm(0,0)$ individually require regularization
-- see the discussion around \eqn{REG} -- any divergence cancels in the linear
combinations of schematic form $\gm(k,\ldots)+2 \zeta_k \gm(0,\ldots)$ in
\eqns{eqn:gamma1G}{eqn:gamma2G}.

The conversion of eMZVs into $\gm$'s amounts to recursively applying the
differential equation \eqn{eqn:tauder} and casting it into the form \eqn{logq2}.
At each step, an instance of $\GG{k}$
is separated until one has reached eMZVs of the form in
\eqns{eqn:gamma1G}{eqn:gamma2G} exclusively.  After converting those into
$\gm$'s, one reverts the direction and successively integrates using
\eqn{eqn:defEInew}, supplementing integration constants from \eqn{eqn:genconst}.

Let us demonstrate the conversion into iterated Eisenstein integrals $\gm$ for
$\om(0,3,5)$.  Employing \eqn{eqn:tauder}, one finds
\begin{equation}
  4\pi^2\frac{\dd }{\dlog q}\omm(0,3,5) = -15 \GG{4}\omm(0,5)+42\omm(0,9)+3\omm(4,5)\,,
  \label{eqn:der1G}
\end{equation}
i.e.~we have $\xi_{4}(0,3,5)= -15 \omm(0,5)$ and 
$\xi_0(0,3,5)= - 42 \omm(0,9) - 3 \omm(4,5)$ 
in the notation of \eqn{logq2}.  While
$\omm(0,5)$ and $\omm(0,9)$ can be readily converted into $\gm$'s using
\eqns{eqn:gamma1G}{eqn:gamma2G}, we will have to take another
derivative\footnote{Alternatively, one could use \eqn{mine7}, but for
illustrational purposes we will perform the recursion explicitly.} for
$\omm(4,5)$:
\begin{align}
  4\pi^2\frac{\dd }{\dlog q}\omm(4,5) 
  & = 9\GG{10}\omm(0)-15\GG{4}\omm(6)+42\omm(10)
  \nnl
  & =9 \GG{10} + 30  \zm_6 \GG{4} +84 \zm_{10}\GG{0}\,.
\end{align}
Performing the integration \eqn{eqn:defEInew} then leads to
\begin{equation}
  \omm(4,5)=9\gm(10) + 30 \zm_6\gm(4) + 84 \zm_{10}\gm(0)\,,
\end{equation}
which -- after plugged into \eqn{eqn:der1G} -- yields
\begin{equation}
  4\pi^2\frac{\dd }{\dlog q}\omm(0,3,5) = -75\GG{4}(\gm(6)+2\zm_{6}\gm(0))+405\gm(10)+90\zm_6\gm(4)+1008\zm_{10}\gm(0)\,.
  \label{eqn:der22}
\end{equation}
After a last integration of the type in \eqn{eqn:defEInew} one finally obtains
\begin{align}
  \omm(0,3,5)&=-405\gm(10,0)-75\gm(6,4) -\zm_6(150\gm(0,4)+90\gm(4,0)) -1008\zm_{10}\gm(0,0)\,,
  \label{eqn:w035result}
\end{align}
which casts the first indecomposable length-three eMZV beyond \eqn{eqn:gamma2G}
into the language of iterated Eisenstein integrals and fits into the pattern
$\gm(k,\ldots)+2 \zeta_k \gm(0,\ldots)$ for the first entries. Further examples
of expressing eMZVs as iterated Eisenstein integrals are listed in
\appref{lesszeros}.


\paragraph{Conversion of weight and length.} Length and weight are different
between the representation of eMZVs in terms of iterated Eisenstein integrals
$\gm$ and the $\omm$-representation. Denoting length and weight for $\gm$ and
$\om$ by $(\ell_\ga,\,w_\ga)$ and $(\ell_\om,\,w_\om)$, respectively, one finds
straightforwardly
\begin{equation}
  \ell_\ga=\ell_\om-1\mand w_\ga=\ell_\om-1+w_\om=\ell_\ga+w_\om \, ,
  \label{eqn:dimshiftG}
\end{equation}
such that
\begin{subequations}
\begin{align}
\gm(k_1,k_2,\ldots,k_n) &\leftrightarrow \te{eMZV in $\omm$-rep. with $\ell_\om=n+1$ and $w_\om=-n+\sum_{j=1}^n k_j$ } 
\label{eqn:translation1G}
\\
\omm(n_1,n_2,\ldots,n_r) &\leftrightarrow \te{Eisenstein integral with $\ell_\ga=r-1$ and $w_\ga=r-1+\sum_{j=1}^r n_j$ }\,. 
\end{align}
\end{subequations}
Those formul\ae{}, however, are valid for the \textit{maximal component} only: as
illustrated e.g.~in \eqn{eqn:gamma2G}, the presentation of eMZVs in terms of
iterated Eisenstein integrals involves different lengths $\ell_\gamma$ and
weights $w_\ga$.  Correspondingly, the maximal component is defined to be comprised from all
terms in an eMZVs $\gm$-representation, which are of length $\ell_\gamma$ and
weight $w_\om$. Below, we will exclude $\gm$'s, which can be represented as
shuffle products, from the maximal component.  Iterated Eisenstein integrals of
length $\ell_\gamma-2 , \ell_\gamma-4,\ldots$ as well as any terms in which
weight is carried by MZVs do not belong to the maximal component. 

The examples in \eqn{eqn:gamma2G} and \eqn{eqn:w035result} give rise to maximal components
\begin{align}
\omm(0,0,n)&=-n(n+1) \gm(n+2,0)+ \te{non-maximal terms} \label{maxcomp1} \\
\omm(0,3,5)&=-405\gm(10,0)-75\gm(6,4) + \te{non-maximal terms}  \ , \label{maxcomp2} 
\end{align}
which are defined up to shuffle products of lower-length iterated Eisenstein
integrals.

Considering \eqn{eqn:translation1G}, one can create $\gm$'s corresponding to
$\om$-representations of negative weight. Since weighting functions $f^{(m)}$
are not defined for negative weight, $\gm(k_1,k_2,\ldots,k_n)$ with
$\sum_{j=1}^n k_j < n$ are clearly incompatible with the definition of eMZVs in
\eqn{l2.7}.  However, the connection with the derivation algebra $\DAlg$ in
\subsecref{subsec:deralg} below will assign a meaning to those $\gm$'s in the
context of relations between eMZVs at length $\ell_\omega \geq 6$.
%

\paragraph{Counting of indecomposable eMZVs.} What are the advantages of
translating eMZVs into iterated Eisenstein integrals? We would like to derive
the set of indecomposable eMZVs with given length and weight from purely
combinatorial considerations, similar to writing down all non-commutative words
of letters $f$ for standard MZVs (cf.~\eqn{UBasis}). In particular, each
indecomposable eMZV in \tabref{tab:emzvbasis} should be related to a particular
combination of shuffle-independent $\gm$'s. Correspondingly, the counting of
indecomposable $\gm$'s of appropriate weight and length should be related to
the numbers in \tabref{tab:basisdim}.

In order to assess the viability of iterated Eisenstein integrals $\gm$ for
this purpose, it is worthwhile to recall the following observations:
\begin{itemize}
 \item[(a)] By construction, constant terms are absent in the
   differential \eqn{eqn:tauder} for eMZVs. This interlocks the first entries
   of iterated Eisenstein integrals representing eMZVs in rigid combinations
   of $\gm(k,\ldots)+2 \zeta_k\gm(0,\ldots)$.  Hence, it is sufficient for
   counting purposes to focus on $\gm(k_1,k_2,\ldots,k_r)$ with $k_1 \neq 0$.
  \item[(b)] The choice of indecomposable eMZVs in \tabref{tab:emzvbasis} contains
    no further divergent representative besides
    $\omm(0,1)=\gm(2)+2\zm_2\gm(0)+\frac{i\pi}{2}$. For any weight and length considered,
    divergences in eMZVs are captured by products with $\gm(2)$ instead of
    shuffle-irreducible integrals of higher length such as
    $\gm(2,4)$. We will assume the continuation of this pattern and confine
    the choice of labels for all other Eisenstein integrals $\gm$ at length
    $\ell_\ga \geq 2$ to the set $\{0,4,6,\ldots\}$. This will be justified
    later on by the observation that the element $\ep_2 \in\DAlg$ corresponding to
    $\gm(2)$ is central by \eqn{poll1}. 
  \item[(c)] The shuffle relations \eqn{gammash} allow to reduce various linear
    combinations of iterated Eisenstein integrals to lower length, e.g.
    \begin{equation}
      \gm(4,4) = \frac{1}{2}\gm(4)^2 \quad\te{and}\quad \gm(6)\gm(4)=\gm(4,6)+\gm(6,4)\,,
      \label{eqn:gammashuffle}
    \end{equation} 
    and the bookkeeping of indecomposable eMZVs boils down to classifying
    shuffle-independent Eisenstein integrals $\gm$. At length $\ell_\ga=2$ and
    weight $w_\ga=10$, possible indecomposable elements read $\gm(10,0)$ and
    $\gm(6,4)$, because $\gm(4,6)$ can be obtained using shuffling of
    $\gm$'s of lower length. Similarly, $\ell_\ga=2$ and $w_\ga=12$ leaves
    no indecomposable eMZVs beyond $\gm(12,0)$ and $\gm(8,4)$.
\end{itemize}
Let us compare the survey of available Eisenstein integrals with the
indecomposable eMZVs in \tabref{tab:emzvbasis}. Eisenstein integrals of length
one immediately match with the maximal component of indecomposable
eMZVs $\omm(0,2n+1)$ of length two using \eqn{eqn:gamma1G}, so the first
non-trivial tests occur at length $\ell_\om=3$, i.e.~$\ell_\ga=2$. 

Via \eqn{eqn:gamma2G} one finds indeed $\gm(4,0),\gm(6,0)$ and $\gm(8,0)$ to
represent the maximal component of $\omm(0,0,2),\omm(0,0,4)$ and $\omm(0,0,6)$,
respectively. For $w_\ga=10$, which corresponds to $w_\om=8$, one can write down
two distinct indecomposable elements: $\gm(10,0)$ and $\gm(6,4)$. This nicely
ties in with the appearance of the second indecomposable eMZV $\omm(0,3,5)$ at 
$\ell_\om=3$ and $w_\om = 8$, see \eqns{maxcomp1}{maxcomp2}.

Similarly, the aforementioned indecomposable eMZVs $\gm(12,0)$ and $\gm(8,4)$
at weight $w_\ga=12$ are in concordance with the $w_\om=10$ entry of
\tabref{tab:emzvbasis},
\begin{align}
  \omm(0,0,10) &= -\frac{\zm_{10}}{3} -110\gm(12,0) -220\zm_{12}\gm(0,0)\nnl
  \omm(0,3,7) &= 
  -294\gm(8,4)-1848\gm(12,0) + \te{non-maximal terms}\ .
  \label{w10l3}
\end{align}
The appearance of the indecomposable eMZVs $\omm(0,3,5)$ and $\omm(0,3,7)$
beyond $\omm(0,0,2n)$ matches the existence of shuffle-independent Eisenstein
integrals $\gm(6,4)$ and $\gm(8,4)$ in addition to $\gm(10,0)$ and
$\gm(12,0)$.
%

\paragraph{Surprises from weight-twelve eMZVs and beyond.}
The literal application of the above reasoning to iterated Eisenstein integrals of 
weight $w_\ga=14$ suggests indecomposable eMZVs
\begin{equation}
\omm(0,0,12) \ \te{from} \ \gm(14,0),\quad
\omm(0,3,9) \ \te{from} \ \gm(10,4) \quad\text{and}\quad
\omm(0,5,7) \ \te{from} \ \gm(8,6)\,.
\end{equation}
This, however, clashes with the findings noted in \tabref{tab:emzvbasis}: at
$\ell_\om=3$ and $w_\om=12$ we find only \textit{two} indecomposable eMZVs
$\omm(0,0,12)$ and $\omm(0,3,9)$, whereas the above counting of appropriate
interated Eisenstein integrals would suggest \textit{three} indecomposable
eMZVs. In particular, $\omm(0,5,7)$ can be expressed in terms of the two
indecomposable eMZVs as written in \eqn{eqn:weight12examples}.

In order to explain the discrepancy between indecomposable eMZVs and
shuffle-independent iterated Eisenstein integrals, let us inspect the first
instance at $w_\ga=14,\ell_\ga=2$, which corresponds to $w_\om=12,\ell_\om=3$.
The natural candidates for indecomposable eMZVs besides $\omm(0,0,12)$ have the
following $\gm$-representations,
\begin{align}
\omega(0,3,9) =&
-315 \gm(8,6)-729 \gm(10,4)-5616 \gm(14,0) + \te{non-maximal terms}\label{w12l3}\\
\omega(0,5,7) =&
 -490 \gm(8,6)-1134 \gm(10,4)-5642 \gm(14,0) + \te{non-maximal terms}\, , \notag
\end{align}
and the relation \eqn{eqn:weight12examples} for $\omm(0,5,7)$ leaves only
$\omm(0,0,12)$ and $\omm(0,3,9)$ indecomposable. In general, there seem to be
non-obvious restrictions to the Eisenstein integrals $\gm$ appearing in eMZVs,
beyond the observations (a), (b) and (c).  In \tabref{tab:indecEIS}, we have
noted the deviations from the expected pattern at lengths $\ell_\om \leq 5$. 
\begin{table}[htp]
\begin{center}
\setlength\tabcolsep{4pt}
\begin{tabular}{|c||  c|c|c|c|c|  c|c|c|c|c|  c|c|c|c|c|  c|c|c|c|}
  \hline  \diagbox{$\ell_\om$}{$w_\om$} &1 &2 &3 &4 &5 &6 &7 &8 &9 &10 &11 &12 &13 &14 &15 &16 &17 &18 &19 \\ \hline \hline   
   2
&1 & &1 & &1
& &1 & &1 &
&1 & &1 & &1
& &1 &    
&1 \\\hline
3  
& &1 & &1 &
&1 & &2 & &2
& &3$_{\textcolor{red}{2}}$ & &3 &
&4$_{\textcolor{red}{3}}$ &  &4$_{\textcolor{red}{3}}$  
& \\\hline
4
&1 & &1 & &2
& &3 & &4 &
&6$_{\textcolor{red}{5}}$ & &8$_{\textcolor{red}{7}}$ & &10$_{\textcolor{red}{8}}$
& &13$_{\textcolor{red}{10}} $ &
&16$_{\textcolor{red}{12}}$   \\\hline
5
& &1 & &2 &
&4 & &6 & &10$_{\textcolor{red}{9}}$
& &14$_{\textcolor{red}{13}}$ & &21$_{\textcolor{red}{17}}$ &
&28$_{\textcolor{red}{23}}$ & &39$_{\textcolor{red}{30}}$ &  \\\hline
%
\end{tabular}
\caption{Number of indecomposable eMZVs at length $\ell$ according to the
  counting of $\gm(k_1,k_2,\ldots,k_n)$ suggested by the above observations
  (a), (b) and (c).  The black numbers denote the number of shuffle-independent
  $\gm$'s with $k_i=0,4,6,\ldots$ and $k_1\neq 0$ while the red numbers
indicate a deviating number of indecomposable eMZVs found from reflection-,
shuffle- and Fay relations or the $q$-expansion.}
\label{tab:indecEIS}
\end{center}
\end{table}
Interestingly, the Eisenstein
integrals $\gm(8, 6)$ and $\gm(10,4)$ enter \eqn{w12l3} and thus any other
eMZV of the same weight and length in the combination 
\begin{equation}
  35 \gm(8,6)+81\gm(10,4)\,
  \label{eqn:gmpollG}
\end{equation}
exclusively. The above quantity is the first in a series of links to the
derivation algebra $\DAlg$ introduced and discussed in the next subsection.


\subsection[A relation to the derivation algebra $\DAlg$]{A relation to the
derivation algebra \texorpdfstring{$\boldsymbol{\DAlg}$}{u}}
\label{subsec:deralg}
The explanation of the deviating numbers for indecomposable eMZVs compared to
shuffle-independent Eisenstein integrals in the last subsection can be provided
starting from the following differential equation for the KZB associator
$A(q)$ defined in \eqn{nils1} \cite{EnriquezEMZV}:
\begin{equation}
\frac{ \dd }{\dlog q} (e^{i\pi [y,x]}A(q)) = \frac{1}{4\pi^2} \left( \sum_{n=0}^{\infty} (2n-1)\GG{2n}(q)\ep_{2n} \right) (e^{i\pi [y,x]}A(q))\,.
\label{eqn:diffA}
\end{equation}
The Eisenstein series $\GG{2n}$ in \eqn{eqn:diffA} are accompanied by
derivations $\ep_{2n}$ which act on the non-commutative variables $x$ and $y$
in the expansion of $A(q)$ via
\begin{subequations}
  \label{eqn:basicrel}
\begin{alignat}{3}
  \ep_{2n}(x)&=(\ad_x)^{2n}(y) \ ,\qquad&&n\geq0\\
  \ep_{2n}(y)&=[y,(\ad_x)^{2n-1}(y)]+\sum_{1\leq j<n}(-1)^j[(\ad_x)^j(y),(\ad_x)^{2n-1-j}(y)] \ ,\quad&&n>0\\
  \ep_0(y)&=0\,.
\end{alignat}
\end{subequations}
They generate a Lie subalgebra $\DAlg$ of the algebra of all derivations on the
free Lie algebra generated by $x,y$ \cite{EnriquezEllAss,KZB,Hain}. The
relations originating from \eqn{eqn:basicrel} have been studied extensively in
ref.~\cite{Pollack}. Beyond
\begin{equation}
[\ep_{2n},\ep_2]=0 \ , \quad n\geq 0 \ ,
\label{poll1}
\end{equation}
there are several non-obvious relations such as 
\begin{subequations}
\label{eqn:poll}
\begin{align}
0=&\,[\ep_{10},\ep_4]-3\,[\ep_{8},\ep_6]\,,
\label{poll2}\\
0=&\,2\,[\ep_{14},\ep_4] - 7\,[\ep_{12},\ep_6] + 11\,[\ep_{10},\ep_8]\,,
\label{poll3}\\
0=&\,80\,[\ep_{12},[\ep_4,\ep_{0}]] + 16\,[\ep_4,[\ep_{12},\ep_0]] - 250\,[\ep_{10},[\ep_6,\ep_0]] \nnl
&- 125\,[\ep_6,[\ep_{10},\ep_0]] + 280\,[\ep_8,[\ep_8,\ep_0]]- 462\,[\ep_4,[\ep_4,\ep_8]] - 1725\,[\ep_6,[\ep_6,\ep_4]]\,.
\label{poll4}
\end{align}
\end{subequations}
The r\^ole of $\ep_2$ as a central element in \eqn{poll1} is reminiscent of
the above observation (b): any appearance of the non-modular $\GG{2}$ can be
captured by powers of $\gm(2)$. Moreover, a peculiar linear combination of
$\gm(8,6)$ and $\gm(10,4)$ has been observed in \eqn{eqn:gmpollG} to appear in
all eMZVs at $\ell_\om=3$ and $w_\om=12$.  Upon identifying labels in $\gm$
with those of derivations $\ep_{2n}$ as suggested by \eqn{eqn:diffA}, one could
attribute the selection rule on $\gm(8,6)$ and $\gm(10,4)$ to \eqn{poll2}.

This connection will be made more precise in the subsequent. For this purpose,
iterated Eisenstein integrals will be rewritten in terms of non-commutative
letters similar to the ones discussed for usual MZVs in \secref{sec:MZV}. In
particular we are led to an structure reminiscent of the $\phi$-map, which
provided the key to a convenient representation of MZVs in which all
known relations over $\mathbb Q$ are automatically built in.

The rewriting of the $\omm$-representation of eMZVs in terms of non-commutative
letters turns out to mimick the procedure used in order to define the map
$\phi$ in \eqn{Phimap}. Despite the resemblance, however, the definition of the
map $\phi$ depends on the choice of an algebra basis for motivic MZVs, while
the rewriting of eMZVs in terms of non-commutative letters to be described
below is completely canonical.
%

\paragraph{Eisenstein integrals as non-commutative words.}
As a first step to make the connection between eMZVs and the algebra of
derivations manifest, let us translate iterated Eisenstein integrals into words
composed from non-commutative generators $g_0,g_2,g_4,\ldots$\footnote{We are
  grateful to Francis Brown who helped us to understand the language and scope
  of non-commutative words in the context of multiple modular values, in
particular for pointing us to section 12 of ref.~\cite{Brown:mmv}.  },
\begin{equation}
\psi \big[ \gamma(k_1,k_2,\ldots,k_n) \big] \equiv \frac{ g_{k_n} g_{k_{n-1}} \ldots g_{k_2} g_{k_1}}{\prod_{j=1}^n (k_j-1)} \ .
\label{eqn:ellphi}
\end{equation}
Here, we need to assume that the iterated Eisenstein integrals are linearly
independent, and the normalization $g_k/(k-1)$ of the non-commutative alphabet
is suggested by the combinations $(k-1) \GG{k}$ in \eqn{eqn:diffA} and the
factors of $n_i \GG{n_i+1}$ in \eqn{eqn:gamma1G}. 

The non-commutative letters $g_{k}$ are naturally endowed with a shuffle
product. The $\psi$-map defined by \eqn{eqn:ellphi} then satisfies
\begin{equation}
\psi\big[ \gamma(n_1,n_2,\ldots,n_r) \gamma(k_1,k_2,\ldots,k_s)
\big] = 
\psi\big[ \gamma(n_1,n_2,\ldots,n_r)
\big] \shuffle
\psi\big[  \gamma(k_1,k_2,\ldots,k_s)
\big]  \ .
\label{gshuffle}
\end{equation}
The linear combination of $\gm(8,6)$ and $\gm(10,4)$ appearing in the eMZVs
with $w_\om=12$ and $\ell_\om=3$ are mapped to
\begin{align}
&\psi \big[ 35\gm(8,6) + 81 \gm(10,4)\big]=
g_6g_8+3 g_4 g_{10} \,.
\label{anderegleichung}
\end{align} 
Hence, the image of any $w_\om=12,\ell_\om=3$ eMZV under \eqn{eqn:ellphi} is
annihilated by the differential operator
\begin{equation}
[\pd_{10},\pd_4] - 3 [\pd_8,\pd_6]\, ,
\label{eqn:pollrln1}
\end{equation}
once differentiation of a non-commutative word in $g_i$ is defined to act on
the leftmost letter
\begin{equation}
\pd_j g_{k_1}\ldots g_{k_n} = \delta_{j,k_1} g_{k_2}\ldots g_{k_n} \,.
\label{proj}
\end{equation}
This differentiation rule satisfies a Leibniz property w.r.t.~the shuffle
product \eqn{gshuffle} and appeared already in the context of the
representation of motivic MZVs in terms of non-commutative letters $f_i$
\cite{Brown:2011ik}, see the discussion in \secref{sec:MZV}.  Note furthermore
that the recursive construction of the eMZVs' $\psi$-image via \eqn{logq2} with
coefficients $\xi_{2k}(n_1,\ldots,n_r)$ determined by the differential equation
(\ref{eqn:tauder}) is similar to the recursive evaluation of the $\phi$-map
\cite{Brown:2011ik}: The coefficients $\xi_{2k+1}$ of $\phi(\zeta^\fm)=
\sum_{3\leq 2k+1 \leq w} f_{2k+1} \xi_{2k+1}$ for some motivic MZV of weight
$w$ are determined by the component of weight $(2k+1) \otimes (w-2k-1)$ in the
coaction. Hence, the $\tau$-derivative in the form \eqn{logq2} exhibits a formal
similarity to the coaction of motivic MZVs. 

However, there is an important difference between the $\phi$-map and the
rewriting of eMZVs in its $\psi$-image: while the $\phi$-map depends on a
choice of algebra generators (for example the adaptation of \eqn{phiExamples}
to the basis in \tabref{zetaBasis}), the $\psi$-map
for eMZVs is completely canonical.

In summary, the $\psi$-image of an eMZV $\omm(n_1,\ldots,n_r)$ is computed in
two steps:
\begin{itemize}
\item use the differential equation to write $\omm(n_1,\ldots,n_r)$ as a linear
  combination of iterated Eisenstein integrals $\gamma(k_1,\ldots,k_s)$.
  Relying on our working hypothesis that iterated Eisenstein integrals are
  linearly independent, this decomposition is unique. 
\item apply the map in \eqn{eqn:ellphi} to each of the $\gamma$'s.
\end{itemize}


\paragraph{Non-commutative differentiation and the derivation algebra $\DAlg$.}
The similarity between \eqns{poll2}{eqn:pollrln1} suggests to identify
derivations $\ep_{2m}$ with derivatives with respect to the non-commutative
letters $\pd_{2m}$. Indeed, we will verify in three steps that the derivations
$\ep_{2m}$ encode the action of $\partial_{2m}$ on the $\psi$-image of the KZB
associator \eqn{nils1} and therefore on the $\psi$-image of any eMZVs:
\begin{itemize}
\item[(i)] integrate the differential equation (\ref{eqn:diffA}) of the KZB
  associator,
  \begin{equation}
  e^{i\pi [y,x]}(A(q) - A(0) ) = e^{i\pi [y,x]} \frac{1}{4\pi^2}  \sum_{n=0}^{\infty} (2n-1) \int_0^q  \dlog q' \  \GG{2n}(q')\ep_{2n} A(q') \ ,
  \label{derprf1}
  \end{equation}
using the corollary $\ep_{2n}([y,x]) = 0$ of \eqn{eqn:basicrel} to commute $\ep_{2n} e^{i\pi [y,x]} = e^{i\pi [y,x]} \ep_{2n}$
\item[(ii)]  apply the $\psi$-map defined in \eqn{eqn:ellphi}: 
  %
  \begin{equation}
  \psi \big[ A(q)-A(0)  \big] =  \sum_{n=0}^{\infty} \ep_{2n} g_{2n} \psi \big[ A(q) \big] \ ,
  \label{derprf2}
  \end{equation}
using the fact that integration against $\frac{ (2n-1) }{4\pi^2} \GG{2n}$
amounts to left-concatenation with $g_{2n}$
\item[(iii)] act with $\partial_{2m}$ such that the sum over $n$ collapses by
  \eqn{proj},
  \begin{equation}
  \partial_{2m}\psi \big[ A(q) - A(0) \big] = \partial_{2m}\psi \big[ A(q) \big] = \partial_{2m} \sum_{n=0}^{\infty} \ep_{2n} g_{2n} \psi \big[ A(q)  \big] =  \ep_{2m}  \psi \big[ A(q)  \big]  \,,
  \label{derprf3}
  \end{equation}
  where we used that the derivative $\pd_{2m}$ annihilates the boundary term
  $A(0)$, which translates into an empty word in the letters $g$. 
\end{itemize}
This is the reason, why any relation among the derivations $\ep_i$ defines a
differential operator via $\ep_i \rightarrow \pd_i$ which annihilates the
$\psi$-image of any eMZV. Explicitly: 
\begin{equation}
\forall \ E\in \mathfrak{u} \ \te{such that} \ E(x) = E(y)= 0 \ \ \ \Rightarrow \ \ \ E \big|_{\epsilon_{2m} \rightarrow \partial_{2m} } \psi\big[ \omega(n_1,\ldots,n_r) \big] = 0 \ .
\label{theorem}
\end{equation}
Thus, any relation in $\mathfrak{u}$ obstructs the appearance of some single linear
combination of iterated Eisenstein integrals \eqn{eqn:defEInew} among eMZVs and
reduces the counting of indecomposable representatives at lengths and weights
governed by the conversion rules \eqn{eqn:translation1G}.


\subsection{Systematics of relations in the derivation algebra}
\label{ssec:systematics}

Naturally, we have been checking the implications of counting
shuffle-independent $\gm(k_1,k_2,\ldots,k_n)$ subject to $k_1\neq 0$ and
$k_i\neq 2$ (cf.~the three observations around \eqn{eqn:gammashuffle}) and the connection with the
derivation algebra $\DAlg$ established in the previous subsection by comparing
$q$-expansions: up to weights $w_\om=30,18,8,5$ for $\ell_\om=3,4,5,6$ we find
complete agreement. There are, however, no obstructions for repeating the
analysis for eMZVs of higher length, as tested for several low weights at
length 7 and 8. 

Counting relations from the algebra of derivations $\DAlg$ for a given weight
and depth works as follows: we start with an ansatz for a relation $E$ of the
form  
\begin{equation}
0 \stackrel{!}{=} \sum_{\{n_1,n_2,\ldots,n_r\} } \alpha_{n_1,n_2,\ldots,n_r} [[\ldots[[ \partial_{n_1},\partial_{n_2}],\partial_{n_3}],\ldots],\partial_{n_r}] 
  \label{eqn:ansatz}
\end{equation}
with rational fudge coefficients $\alpha_{n_1,n_2,\ldots,n_r}$ and
$\{n_1,n_2,\ldots,n_r\}$ composed of $n_i=0,4,6,\ldots$ of appropriate weight
and length.  The number $r$ of partial derivatives in the nested commutators of
\eqn{eqn:ansatz} (or the number of $\ep_{n}$ in the dual derivations,
respectively) is referred to as \textit{depth}. Of course, the summation in
\eqn{eqn:ansatz} is restricted to nested commutators which are independent
under Jacobi identities. 

Considering \eqn{theorem}, the above ansatz for $E$ should annihilate all
$\psi$-images of eMZVs of the length and weight considered. Using a
sufficiently large set of eMZVs, one can easily fix all fudge coefficients in
the ansatz and thus extract relations.

Using this method, we find perfect agreement of \eqn{eqn:ansatz} as an operator
equation acting on eMZVs with the relations in the derivation algebra available
in refs.~\cite{Pollack,BauSch, BauSch2}. In the following
paragraphs we will review their classification and extend the explicit results
to higher commutator-depths.


\paragraph{Special r\^ole of $\ep_2$.} As already observed above, none of the
indecomposable eMZVs besides $\omm(0,1)$ does contain an Eisenstein integral
involving $\GG{2}$. This reflects the r\^ole of $\ep_2$ as a central element, as
noted in \eqn{poll1}. Hence, it is sufficient to study commutator relations
without $\ep_2$.


\paragraph{Irreducible versus reducible relations.} Any relation in the
derivation algebra $\DAlg \ni E=0$ of the form \eqn{eqn:ansatz} yields an
infinity of higher-depth corollaries by repeated adjoint action of~$\ep_{n}$:
\begin{equation}
E=0 \ \ \ \Rightarrow \ \ \ 
\ad_{n_1,n_2,\ldots,n_k} (E) \equiv [\ep_{n_1},[\ep_{n_2},[\ldots ,[\ep_{n_k},E]\ldots ]]] =0 \ .
\label{surpr7}
\end{equation}  
Any instance of \eqn{surpr7} with $k>0$ and $E$ denoting a vanishing combination of
$\ep_n$-commutators is called a \textit{reducible relation}, whereas relations
that cannot be cast into the form $\ad_{n_1,n_2,\ldots,n_k} (E)=0$ are referred
to as \textit{irreducible}. For instance, the simplest non-obvious relation
\eqn{poll2} is irreducible and gives rise to reducible relations such as
\begin{equation}
 \big[ \ep_{n}, [\ep_{10},\ep_4] - 3 [\ep_8,\ep_6] \big] = 0 \, ,
\end{equation}
and generalizations to higher depth. They affect the bookkeeping of irreducible
eMZVs starting from $w_\ga=14$ and $\ell_\ga=3$, which corresponds to
$w_\om=11$ and $\ell_\om=4$.

A correspondence between cusp forms of weight $w$ and irreducible relations at
depth $d$ and weight $w+2(d-1)$ has been discussed in ref.~\cite{Pollack}. In
the same way as the number of cusp forms at modular weight $w$ is given by
\begin{equation}
\chi_w \equiv  \left\{
\begin{array}{cl} \lfloor \frac{w}{12} \rfloor  -1 &: \ w = 2 \ \te{mod} \ 12 \\
\lfloor \frac{w}{12} \rfloor  &: \ \te{other even values of $w$}
\end{array} \right. \ ,
\label{chicount}
\end{equation}
we expect $\chi_{w-2(d-1)}$ irreducible relations at weight $w$ and depth $d$
relevant to eMZVs of non-negative weight $w_\om$ (see \eqn{eqn:dimshiftG} for
its relation to the weight of the iterated Eisenstein integral). In
\tabref{tab:irreducible}, this conjectural 
counting is exemplified up to $w_\ga=30$
with a notation $r_{w_{\ga}}^{d}$ for such irreducible relations.
\begin{table}[htp]
\begin{center}
  \setlength\tabcolsep{8pt}
  \renewcommand{\arraystretch}{1.2}
\begin{tabular}{|c||>{$}c<{$}|>{$}c<{$}|>{$}c<{$}|>{$}c<{$}|>{$}c<{$}|>{$}c<{$}|>{$}c<{$}|>{$}c<{$}|>{$}c<{$}|}
  \hline
  \diagbox{$w_\ga$}{$\ell_\ga$} &2&3&4&5&6&7&8&9&10
\\\hline\hline
12 &0&0&0&0&0&0&0&0&0
\\\hline
14 &r_{14}^2&0&0&0&0&0&0&0&0
\\\hline
16 &0&r_{16}^3&0&0&0&0&0&0&0
\\\hline
18 &r_{18}^2&0&r_{18}^4&0&0&0&0&0&0
\\\hline
20 &r_{20}^2&r_{20}^3&0&r_{20}^5&0&0&0&0&0
\\\hline
22 &r_{22}^2 &r_{22}^3&r_{22}^4&0&r_{22}^6&0&0&0&0
\\\hline
24 &r_{24}^2&r_{24}^3&r_{24}^4&r_{24}^5&0&r_{24}^7&0&0&0
\\\hline
26 &2\times r_{26}^2&r_{26}^3&r_{26}^4&r_{26}^5&r_{26}^6&0&r_{26}^8&0&0
\\\hline
28 &r_{28}^2&2\times r_{28}^3&r_{28}^4&r_{28}^5&r_{28}^6&r_{28}^7&0&r_{28}^9&0
\\\hline
30 &2\times r_{30}^2&r_{30}^3&2\times r_{30}^4&r_{30}^5&r_{30}^6&r_{30}^7&r_{30}^8&0&r_{30}^{10}
\\\hline
\end{tabular}
\caption{Irreducible relations $r_w^\ell$. Up to weight 30 there are no more
  than two relations at a particular weight and length, which will, however,
  change proceeding to higher weight and length. An actual list of the first
  irreducible relations is available in \protect\appref{sec:knownrelations}.}
\label{tab:irreducible}
\end{center}
\end{table}
Relations of depth two can be cast into a closed formula \cite{LNT}
\begin{align}
0 &= \sum_{i=1}^{2n+2p-1} \frac{ [\ep_{2p+2n-i+1},\ep_{i+1}] }{(2p+2n-i-1)!}
\Big\{ \frac{ (2n-1)! B_{i-2p+1} }{(i-2p+1)!}+\frac{ (2p-1)! B_{i-2n+1} }{(i-2n+1)!} \Big\}\ ,
\label{closedpoll}
\end{align}
where $p,n\geq 1$ denote arbitrary integers and $B_n$ are Bernoulli numbers.
Each term of \eqn{closedpoll} carries weight $2(p+n+1)$, e.g.~the weight-14
relation \eqn{poll2} follows from any partition of $p+n=6$, and the
weight-18 relation \eqn{poll3} from any partition of $p+n=8$.

Irreducible relations at higher depth can be obtained in electronic form from
the website \EMZVDatamine{}, whereas relations of depth three at $w=16, 20$ and
depth four at $w=18, 22$ are provided in ref.~\cite{Pollack}.  New relations
beyond those in said reference are obtained from the differential operators
\eqn{eqn:ansatz} annihilating all eMZVs of corresponding weight and length.
This approach to finding relations in the derivation algebra appears
computationally more efficient to us than evaluating the action of elements of
the derivation algebra on generators $x$ and $y$ of the free Lie algebra.
However, once a candidate relation has been identified, it is straightforward
to check its validity using its action on the letters $x$ and $y$ via
\eqn{eqn:basicrel}. 


\paragraph{Vanishing nested commutators.} Starting from $w_\ga=8$ and
$\ell_\ga=5$, we find that the $\psi$-image of any eMZV with appropriate weight
and length is annihilated by operators of the form
\begin{equation}
  [[[[\partial_4,\partial_0],\partial_0],\partial_0],\partial_{2m}] \ .
  \label{eqn:surprise1}
\end{equation}
The reason becomes clear by considering $\gm(4,0,0,0)$, one of the
corresponding Eisenstein integrals. By \eqn{eqn:dimshiftG}, related eMZVs are
bound to have $\ell_\om=5$ and $w_\om=0$, but the only eMZV with these
properties is $\omm(0,0,0,0,0)=1/120$ which cannot equal the non-constant
$\gm(4,0,0,0)$.  Hence, the latter does not occur among eMZVs and signals the
irreducible relation
\begin{equation}
  [[[\ep_4,\ep_0],\ep_0],\ep_0]=0 \ ,
  \label{eqn:surprise1a}
\end{equation}
which in turn implies that $[[[\partial_4,\partial_0],\partial_0],\partial_0]$
annihilates the KZB associator by \eqn{derprf3}. The relation
\eqn{eqn:surprise1a} can be understood from the organization of $\DAlg$ in
terms of representations of the Lie algebra $\mathfrak{sl}_2$: considering $\ep_{2m}$ as the
lowest-weight state in a $(2m-1)$-dimensional module, the highest-weight vector
$\ad^{2m-2}_{0}\ep_{2m}$ is annihilated by further adjoint action of $\ep_0$.

Further irreducible relations of this type include
\begin{equation}
\ad^{p-1}_{0} \ep_p =
\underbrace{[ \ep_0, \ldots[\ep_0, [\ep_0,}_{p-1\ \te{times}}
\ep_p]]\ldots ]
 = 0 \co p = 4,6,8,\ldots \ ,
\label{morenegweight}
\end{equation}
corresponding to the Eisenstein integral $\gm(p,0^{p-1})$ with would-be eMZV
partners of vanishing $w_\om$. Different partitions of the weight in
\eqn{morenegweight} lead to further relations such as 
\begin{equation}
[[[[[[[\ep_4,\ep_0],\ep_4],\ep_0] ,\ep_0],\ep_0],\ep_0],\ep_0] =0 \co [[[[[[[[[\ep_4,\ep_0],\ep_6],\ep_0] ,\ep_0],\ep_0],\ep_0],\ep_0] ,\ep_0],\ep_0] = 0 \ .
\label{evenmorenegweight}
\end{equation}
Since all their permutations via $\ep_4\leftrightarrow \ep_0$ or
$\ep_6\leftrightarrow \ep_0$ can be identified as a reducible relation
descending from \eqn{morenegweight}, we expect no further irreducible relations
at $d=w_\gamma=8$ or $10$ besides \eqn{evenmorenegweight}.


\paragraph{Additional generators of the Lie algebra} Consider the free Lie
algebra $\mathfrak{k}=\ZL(z_3,z_5,z_7,\ldots)$ generated by one element in
every odd degree strictly greater than one. As mentioned on page 6 of
ref.~\cite{Pollack}, every generator $z_{2k+1}$ of $\mathfrak{k}$ defines a
derivation $\tdz_{2k+1}$ of depth $2k+1$ and weight $4k+2$ of the free Lie algebra on two
generators $x,y$, and satisfies $[\tdz_{2k+1},\DAlg] \subset \DAlg$.
More precisely, the elements $\ep_0,\ep_2$ are
annihilated by the elements $\tdz_{2k+1}$
\begin{equation}
0 = [\tdz_{2k+1},\ep_0] = [\tdz_{2k+1},\ep_2] \co k=1,2,3,\ldots \ ,
\label{surpr2}
\end{equation}
and their commutators with $\ep_4,\ep_6,\ldots$
can be constructed using the techniques of \cite{Pollack}, e.g.~
\begin{equation}
 [\tdz_{3},\ep_4]=
 -\frac{1}{14} [ [ \epsilon _4,\epsilon _0],[ \epsilon _6,\epsilon _0]]+\frac{1}{42} [ \epsilon _4,[ \epsilon _0,[ \epsilon _0,\epsilon _6]]]-\frac{1}{7} [ \epsilon _6,[ \epsilon _0,[ \epsilon _0,\epsilon _4]]]\,.
  \label{simpletz}
\end{equation}
They give rise to further reducible relations, starting from length five at
weights $20,24,26,\ldots$ by the commutator of $\tdz_3$ with the depth-two
relations in \eqn{eqn:poll} or \eqn{closedpoll}.


\subsection{Counting relations between nested commutators}


\paragraph{Example.} In order to demonstrate the virtue of the derivation
algebra as a counting formalism for indecomposable eMZVs, let us consider
$w_\ga=20, \ell_\ga=5$ as a specific example, which corresponds to $w_\om=15,
\ell_\om=6$. This is the first situation, where all four types of relations
described in the previous section have to be taken into account in order to
arrive at what we believe is the correct counting of eMZVs.

The na\"ive enumeration of shuffle-independent $\gm$'s with $k_1\neq 0$ and
$k_i \neq 2$ leads to 55 distinct elements.  Each relation of depth 5 and
weight 20 in the derivation algebra will lower this number according to
\eqn{theorem}. 

Let us first consider reducible relations. Starting from
\tabref{tab:irreducible}, one can construct the following reducible relations
by adjoint action of $\ep_{n}$ (recalling the notation $r^d_{w_\ga}$ for
irreducible relations of depth $d$ and weight $w_\ga$ as well as
$\ad_{n_1,n_2,\ldots,n_k} r_i^j \equiv [\ep_{n_1},[\ep_{n_2},[\ldots
  ,[\ep_{n_k},r^d_{w_\ga}]\ldots ]]]$):
\begin{equation}
\label{eqn:ex1}
\begin{array}{rlcccrl}
  \ad_{6,0,0}\,r_{14}^2&\leftrightarrow \ \ \, \te{3 permutations} \ , &&&  &\ad_{0,0,0}\,r_{20}^2&\leftrightarrow \ \ \,\te{1 permutation}\\
  \ad_{4,0}\,r_{16}^3&\leftrightarrow \ \ \,\te{2 permutations}  \ , &&& &  \ad_{0,0}\,r_{20}^3&\leftrightarrow \ \ \, \te{1 permutation}\,
\end{array}\ .
\end{equation}
In addition, there is one relation each descending from the vanishing nested
commutator \eqn{eqn:surprise1a} and the additional Lie algebra generator
$\tdz_3$,
\begin{equation}
  [[[[\ep_4,\ep_0],\ep_0],\ep_0],\ep_{16}]=0\quad\te{and}\quad[\tdz_3, r_{14}^2]=0\,,
  \label{eqn:ex2}
\end{equation}
which makes a total of 9 reducible relations. 

Indeed, starting with an ansatz of the form \eqn{eqn:ansatz}, we find ten
distinct relations: while \eqns{eqn:ex1}{eqn:ex2} are confirmed, our method
explicitly delivers the new irreducible relation $r_{20}^5$ expected from
\tabref{tab:irreducible}.  To our knowledge this is the first appearance of an explicit
relation at depth ~5 in $\DAlg$, which is written out in \appref{eqn:newlength5}.
Correspondingly, we find the number of indecomposable eMZVs at
$(\ell_\gamma,w_\gamma)=(5,20)$ (or $(\ell_\om,w_\om)=(6,15)$) to be 45. 


\paragraph{General.} In order to repeat the counting procedure from the above
example for a variety of weights and lengths, the following tables give an
overview of the required ingredients: The numbers of shuffle-independent
iterated Eisenstein integrals compatible with observations (a) and (b) in
subsection \ref{ssec:eisint} are gathered in table \ref{tab:gamcount} and have
to be compared with the counting of relations in $\DAlg$ seen in table
\ref{tab:allPollack}. 
\begin{table}[htp]
\begin{center}
\setlength\tabcolsep{4pt}
\begin{tabular}{|c||  c|c|c|c|c|  c|c|c|c|c|  c|c|c|c|c|  c|
  }
  \hline  \diagbox{$\!\ell_\gamma \!\! \!\!\!\!\! \!\!$}{$\!\! \!w_\gamma\!$} &0 &2 &4 &6 &8 
  &10 &12 &14 &16 &18 
  &20 &22 &24 &26 &28 
  &30 
  \\ \hline \hline 
  1 &1 &1&1&1&1&1  &1&1&1&1&1 &1&1&1&1&1 
  \\\hline
2 &0& 0& 1& 1& 1& 2& 2& 3& 3& 4&
       4& 5& 5& 6& 6& 7 
       \\\hline
3& 0& 0& 1& 1& 2& 3& 4& 6& 8& 
      10& 13& 16& 19& 23& 27& 31
      \\\hline
4 & 0& 0& 1& 1& 2& 4& 6& 10& 14& 
      21& 28& 39& 50& 66& 82& 104
      \\\hline
 5 &0& 0& 1& 1& 3& 5& 9& 15& 24& 
      37& 55& 80& 113& 156& 211& 280
      \\\hline
6&       0& 0& 1& 1& 3& 6& 11& 21& 35& 
      59& 93& 146& 217& 322& 459& 649
      \\\hline
7& 0& 0& 1& 1& 4& 7& 15& 28& 51& 
      89& 150& 245& 389& 602& 910& \!1347\!
      \\\hline
\end{tabular}
\caption{Shuffle-independent $\gm(k_1,\ldots,k_n)$ subject to $k_1\neq 0$ and $k_i \neq 2$ at various weights $w_\gamma$ and lengths $\ell_\gamma$.}
\label{tab:gamcount}
\end{center}
\end{table}
\begin{table}[htp]
\begin{center}
\setlength\tabcolsep{4pt}
\begin{tabular}{|c||  c|c|c|c|  c|c|c|c|c|  c|c|c|c|c|  c|c|c|c|c| c|}
  \hline  \diagbox{$\!d \!\! \!\!\!\!\! \!\!$}{$\!\! \!w_\gamma\!$}  &2 &4 &6 &8 
  &10 &12 &14 &16 &18 
  &20 &22 &24 &26 &28 
  &30 &32 &34 &36 &38 
  &40
  \\ \hline \hline 
2   
 &0& 0& 0& 0& 0& 0& 1& 0& 1& 1& 1&1 & 2& 1& 2& 2& 2& 2& 3& 2
       \\\hline
3&0&0&0&0&0&0&1&1&2&3 &4 &5 &7 &8 &10 &12 &14 &16 &19 &21
 \\\hline
4 & 0&1&0&0&0&0&1&1&4 &5 &9 &13 &19 &? &? &? &? &? &? &? 
\\\hline
 5 & 0 &1 &0 &1 &1 &1 &2 &2 &6 &10 &? &? &? &? &? &? &? &? &? &?
 \\\hline
6   &0 &1 &1 &2 &2 &3 &5 &6 &11 &? &? &? &? &? &? &? &? &? &? &?
\\\hline
\end{tabular}
\caption{Relations in the derivation algebra at various weights $w_\gamma$ and
depths $d$, excluding the central element $\ep_2$.}
\label{tab:allPollack}
\end{center}
\end{table}
Once the offset between $(w_\gamma,\ell_\gamma)$ and $(w_\omega,\ell_\omega)$
in \eqn{eqn:translation1G} is taken into account, one arrives at the numbers of
indecomposable eMZVs in the $\omm$-representation noted in
\tabref{tab:emzvbasisdim}. 

\begin{table}[htp]
\begin{center}
\setlength\tabcolsep{4pt}
\begin{tabular}{|c||  c|c|c|c|c|  c|c|c|c|c|  c|c|c|c|c|  c|c|c|c| c|c|c|c|}
  \hline  \diagbox{$\ell_\omega$}{$w_\omega$} &1 &2 &3 &4 &5 &6 &7 &8 &9 &10 &11 &12 &13 &14 &15 &16 &17 &18 &19 &20 &21 &22 &23\\ \hline \hline   
   2
&1 & &1 & &1
& &1 & &1 &
&1 & &1 & &1
& &1 &    
&1 
& &1 & &1
\\\hline
3  
& &1 & &1 &
&1 & &2 & &2
& &2 & &3 &
&3 &  &3  
& &4 & &4
& \\\hline
4
&1 & &1 & &2
& &3 & &4 &
&5 & &7 & &8
& &10 & &\textcolor{blue}{12}   
& &\textcolor{blue}{14} & &\textcolor{blue}{16}

\\\hline
5
& &1 & &2 &
&4 & &6 & &9
& &13 & &\textcolor{blue}{17} &
&\textcolor{blue}{23} & &\textcolor{blue}{30} &  
&\textcolor{blue}{37} & &\textcolor{blue}{47} &
\\\hline
6  
&1 & &2 & &4
& &8 & &13 &
&\textcolor{blue}{22} & &\textcolor{blue}{31} & &\textcolor{blue}{45}
& &\textcolor{blue}{?}  &  &\textcolor{blue}{?}
& &\textcolor{blue}{?} & &\textcolor{blue}{?}
\\\hline
7 
& &1 & &\textcolor{blue}{4} &
&\textcolor{blue}{8} & &\textcolor{blue}{16} & &\textcolor{blue}{29}
& &\textcolor{blue}{48} & &\textcolor{blue}{?} &
&\textcolor{blue}{?} &  &\textcolor{blue}{?} & &\textcolor{blue}{?} & &\textcolor{blue}{?} &
\\\hline
\end{tabular}
\caption{Numbers $N(\ell_\om,w_\om)$ of indecomposable eMZVs in their
  $\om$-representation. This is an extended version of
  \protect\tabref{tab:basisdim}, where the black results are obtained by
  explicitly determining $q$-expansions while results printed in blue originate
  from testing relations between nested commutators as described around
  \protect\eqn{eqn:ansatz}.
  }
\label{tab:emzvbasisdim}
\end{center}
\end{table}
From the above data, one readily arrives at all-weight statements on the number
of indecomposable eMZVs of length $\ell_\om \leq 4$:
\begin{itemize}
\item At length $\ell_\om = 2$, there is obviously one indecomposable eMZV at
  each odd weight $w_\om$.
\item At length $\ell_\om = 3$, the number of indecomposable eMZVs at even
  weight $w_\om$ is $\lceil \frac{1}{6}w_\om\rceil$. This follows from
  comparing the number $\left \lceil \frac{ w_{\om} }{4} \right \rceil - 1$ of
  admissible $\gamma(k_1,k_2)$ ($k_1 >k_2, \ k_i \neq 2$) at weight $w_{\om} >
  4$ with the counting of depth-two relations in $\DAlg$ governed by
  \eqn{chicount}.
\item At length $\ell_\om = 4$, the number of indecomposable eMZVs at odd
  weight $w_\om$ is conjectured to be $\lfloor \frac{1}{2}+
  \frac{1}{48}(w_\om+5)^2 \rfloor$. This conjecture stems from extrapolating
  \cite{oeis} the data available at $w_\om\leq 37$. The extrapolation will
  remain valid, if the counting of irreducible $r^3_{w}$ keeps on following the
  cusp forms. 
\end{itemize}
Starting from the next length, $\ell_\om=5$ or $\ell_\ga=4$, an effect
well-known from the algebra of MZVs kicks in: because the lowest non-trivial
relation from the derivation algebra $\DAlg$ exists at weight 14 depth 2, there
is the possibility to obtain the ``relation of a relation''
$\te{ad}_{r^2_{14}}(r^2_{14})=0$ at weight 28, depth 4. This effect, which
appears in iterated form for higher depth, as well as the action of the
generators of the free Lie algebra $\mathfrak{k}$ described in
\subsecref{ssec:systematics} render the counting at higher depth difficult.
Correspondingly, a closed formula, e.g.~a generating series for the number of
indecomposable eMZVs at given length and weight is still lacking and some of the
entries in table \ref{tab:allPollack} are left undetermined.


\subsection{A simpler representation of the eMZV subspace}
\label{sec:simple}

From the discussion in the previous subsections it became clear that eMZVs can
be nicely represented in terms of iterated Eisenstein integrals
\eqn{eqn:defEInew}. While those integrals have to be regularized individually
as pointed out in the context of \eqn{REG}, the representation of eMZVs cannot
involve any divergences upon integrating their $\tau$-derivative
\eqn{eqn:tauder}. In this section we would like to manifest this property and
define a modified version of iterated Eisenstein integrals $\gmz$, which are
individually convergent by construction. By using the $\gmz$-language, one will
trade some of the connections to periods and motives \cite{Brown:mmv} inherent
in the $\gm$-language for compactness of representation. A further advantage of
the $\gmz$-language to be introduced is a better accessibility of the
$q$-expansions of eMZVs. 


\paragraph{Modified iterated Eisenstein integrals.} Already in
\subsecref{ssec:eisint} it was remarked that the $\tau$-derivative of eMZVs
determined by the differential equation (\ref{eqn:tauder}) cannot contain any
constant terms. Therefore, it is an obvious idea to subtract the constants from
the non-trivial Eisenstein series before defining their iterated integrals:
\begin{align}
 \GGn0 &\equiv -1
\notag \\
\GGn{k} &\equiv \GG{k} - 2\zm_k = \frac{2(-1)^{k/2}(2\pi)^k}{(k-1)!} \sum_{m,n=1}^{\infty} m^{k-1} q^{mn}
  \co \ k \ \te{even}, \ k \neq 0  \,.
  \label{eqn:defG0}
\end{align}
Using this definition, one can rewrite \eqns{eqn:tau3}{eqn:tau4} as
\begin{subequations}
\begin{alignat}{3}
  \frac{\dd}{\dlog q} \omm(0,n) &= \frac{n}{4\pi^2} \GGn{n+1},\qquad &&n \ \te{odd}
\label{eqn:tau32G0} \\
\frac{\dd}{\dlog q} \omm(0,0,n) &= \frac{n}{4\pi^2} \omm(0,n+1)\GGn0,\qquad &&n \ \te{even}  \ ,
\label{eqn:tau42G0}
\end{alignat}
\end{subequations}
and the differential equation (\ref{eqn:tauder}) for generic eMZVs can be easily cast into the form
\begin{equation}
  \frac{\dd }{\dlog q} \omm(n_1,n_2,\ldots,n_r)= \frac{1}{4\pi^2} \sum_{k=0}^{\infty} \rho_{2k}(n_1,n_2,\ldots,n_{r}) \GGn{2k}\,.
\label{logq2G0}
\end{equation}
In complete analogy to
\eqn{logq2}, the coefficients $\rho_{2k}(n_1,\ldots,n_{r}) $ are linear combinations of eMZVs with
weight $n_1+\ldots+n_r+1-2k$ and length $r-1$, the only difference being that Eisenstein series in
\eqn{eqn:tauder} are now expanded via $\GG{k}=\GGn{k}-2\zm_k \GGn{0}$ whenever $k \neq 0$.

From the form \eqn{logq2G0} of the differential \eqn{eqn:tauder}, it is straightforward to introduce
modified iterated Eisenstein integrals $\gmz$ via
\begin{align}
&\gmz(k_1,k_2,\ldots,k_n;q) \equiv
\frac{1}{4\pi^2} \int_{0 \leq q'\leq q} \dlog q' \ \gmz(k_1,\ldots,k_{n-1};q') \GGn{k_n}(q') \co k_1\neq 0
\label{eqn:defEI} \nnl
&\ \ \ \ = \frac{1}{(4\pi^2)^n} \int_{0 \leq q_i<q_{i+1} \leq q} \dlog q_1\  \GGn{k_1}(q_1) \ \dlog q_2 \ \GGn{k_2}(q_2) \ \ldots \ \dlog q_n \GGn{k_n}(q_n)  \ ,
\end{align}
whose definition as an iterated integral implies
\begin{align}
\frac{ \dd }{\dlog q} \gmz(k_1,k_2,\ldots,k_n;q) &= \frac{ \GGn{k_n}(q) }{4\pi^2} \gmz(k_1,k_2,\ldots,k_{n-1};q)
\label{eqn:diffgam2} \\
\gmz(n_1,n_2,\ldots,n_r;q) \gmz(k_1,k_2,\ldots,k_s;q) &= \gmz \!  \big( (n_1,n_2,\ldots,n_r) \shuffle (k_1,k_2,\ldots,k_s) ;q
\big)\,.
\label{gammash2}
\end{align}
The notion of weight and length are not altered w.r.t.~the definition for
$\gm$. Naturally, $\gmz$'s suffer from the same caveat with respect to linear
independence as their cousins $\gm$. There are several advantages of employing
this modified class of iterated Eisenstein integrals $\gmz$ for the description
of eMZVs:
\begin{itemize}
  \item Logarithmic divergences for $q\to 0$ as present in \eqn{eqn:defEInew}
    do not occur after setting $k_1\neq0$. Modified iterated Eisenstein
    integrals $\gmz$ are perfectly well-defined objects which do not require
    regularization.
  \item The number of terms necessary to express eMZVs as combinations of
    iterated Eisenstein integrals $\gmz$ is significantly lower than for $\gm$.
  \item The absence of constant terms in the expansion of $\GGn{k_1}$
    propagates to any convergent iterated Eisenstein integral,
     \begin{equation}
     \gmz(k_1,k_2,\ldots,k_n;0) = 0 \ .
     \label{qzero}
     \end{equation}
\end{itemize}
Note that we will again suppress the dependence on $q$ in most cases:
$\gmz(\ldots)\equiv\gmz(\ldots;q)$\,.


\paragraph{Examples.} Let us return to the examples
\eqns{eqn:tau32G0}{eqn:tau42G0}.  The differential equation
(\ref{eqn:diffgam2}) immediately implies
\begin{subequations}
\begin{alignat}{3}
  \omm(0,n)&=\delta_{1,n}\frac{\pi i}{2}+n\gmz(n+1),\quad&&n \ \te{odd}
\label{eqn:gamma1} \\
 \omm(0,0,n)&=-\frac{1}{3}\zm_n - n(n+1) \gmz(n+2,0),\quad&&n \ \te{even}\,,
\label{eqn:gamma2}
\end{alignat}
\end{subequations}
where $\delta_{1,n}\frac{\pi i}{2}$ and $-\frac{1}{3}\zm_n$ arise as
integration constants w.r.t.~$\log q$.  Indeed, these expressions are
convergent by definition and shorter than their counterparts in
\eqns{eqn:gamma1G}{eqn:gamma2G}.

For illustrational purposes let us also revisit the example $\om(0,3,5)$.
Its derivative
\begin{align}
  4\pi^2\frac{\dd }{\dlog q}\omm(0,3,5) 
& =   30\zm_6\om(0,3) -15 (\GGn{4}-2\zm_4\GGn{0})\omm(0,5) + 45 \omm(0,9)
  \label{eqn:der1}
\end{align}
amounts to $\rho_{4}(0,3,5)= -15 \omm(0,5)$ and $\rho_0(0,3,5)=30 \zm_{4}
\omm(0,5) - 45 \omm(0,9) - 30\zm_6 \omm(0,3)$ in the notation of \eqn{logq2}
and can be translated to modified Eisenstein integrals via \eqn{eqn:gamma1}:
\begin{equation}
  4\pi^2\frac{\dd }{\dlog q}\omm(0,3,5) = 90 \zm_6\gmz(4) -75\,(\GGn{4}-2\zm_4\GGn{0})\gmz(6)+405\gmz(10)
  \label{eqn:der2}
\end{equation}
Integration using \eqn{eqn:defEI} yields the following alternative
representation to \eqn{eqn:w035result},
\begin{equation}
  \omm(0,3,5)=-90 \zm_6\gmz(4,0) + 150 \zm_4\gmz(6,0) - 75 \gmz(6,4) -405\gmz(10,0)\,.
  \label{forapp}
\end{equation}
Further examples of eMZVs expressed in the language of modified iterated
Eisenstein integrals can be found in \appref{app:conv}.


\paragraph{\texorpdfstring{$\boldsymbol{q}$}{q}-expansion.} In contrast to the
$\gm$-language used in the last section, there is no caveat on regularization
when performing the integrals over $q_j$ in the definition \eqn{eqn:defEI} of
$\gmz$.  The $q$-expansion stems from the expression for Eisenstein series in
\eqn{eqn:defG0} and can be cast into a closed form (with $0^{n}$ denoting a
sequence of $n$ entries $0,0,\ldots,0$):
\begin{align}
&\gmz(k_1,0^{p_1-1},k_2,0^{p_2-1},\ldots,k_r,0^{p_r-1}) =
\prod_{j=1}^{r} \Big( - \frac{ 2 (2\pi i)^{k_j - 2 p_j} }{(k_j-1)!} \Big)
\label{qgamma1}\\
& \ \ \ \ \ \ \ \times \sum_{m_i,n_i=1}^{\infty} \frac{m_1^{k_1-1} m_2^{k_2-1} \ldots m_r^{k_r-1}  q^{m_1n_1+m_2n_2+\ldots +m_rn_r}}{(m_1 n_1)^{p_1} (m_1n_1+m_2n_2)^{p_2} \ldots (m_1n_1+m_2n_2+\ldots +m_rn_r)^{p_r}}  \ .
\notag
\end{align}
An even more compact representation can be achieved using the divisor sum 
\begin{equation}
\sigma_k(n) \equiv \sum_{d|n} d^k\,,
\label{qgamma3}
\end{equation}
which allows to rewrite \eqn{qgamma1} as
\begin{align}
&\gmz(k_1,0^{p_1-1},k_2,0^{p_2-1},\ldots,k_r,0^{p_r-1}) =
\prod_{j=1}^{r} \Big( - \frac{ 2 (2\pi i)^{k_j - 2 p_j} }{(k_j-1)!} \Big)
\label{qgamma2} \\
& \ \ \ \ \ \ \ \times \sum_{0<n_1<n_2<\cdots<n_r} \frac{\sigma_{k_1-1}(n_1) \sigma_{k_2-1}(n_2-n_1) \ldots \sigma_{k_r-1}(n_r-n_{r-1}) q^{n_r}} {n_1^{p_1} n_2^{p_2} \ldots n_r^{p_r}} \,.
\notag
\end{align}
The above expression bears some resemblance to the sum representation
\eqn{mzvsum} of MZVs. One could wonder if rearrangements of the sums could
yield a genus-one analogue of stuffle relations. However, both the appearance
of the divisor sums and the $q$-dependence prevent such manipulations.  In
fact, we did not observe a single relation among iterated Eisenstein integrals
$\gmz$ beyond the shuffle relations \eqn{gammash} up to weights $44,31,22,19$
for length $2,3,4,5$, respectively.

Given the above $\gmz$-representation of the simplest eMZVs, we arrive at two
closed forms for $q$-expansions
\begin{align}
\omm(0,k) &=\delta_{k,1} \frac{i\pi}{2} 
+  \frac{2 (-1)^{(k+1)/2} (2\pi)^{k-1} }{(k-1)!}\sum_{m,n=1}^{\infty} \frac{ m^{k-1}}{n} q^{mn} \co k \ \te{odd}
\label{tau3a}
\\
\omm(0,0,k) &= -\frac{1}{3}\zeta_k 
+ \frac{2 (-1)^{(k+2)/2} (2\pi)^{k-2}}{(k-1)!}
 \sum_{m,n=1}^{\infty}  \frac{ m^{k-1}}{n^2} q^{mn} \co k \ \te{even}\,,
\label{tau3b}
\end{align}
while further expressions for interesting $\omm(0,0,\ldots,0,k)$ at higher length
are given in \appref{manyzeros}.


\paragraph{Connection with the derivation algebra.}  A manifestly convergent
description of eMZVs in terms of modified iterated Eisenstein integrals $\gmz$
comes with a price at the end of the day: the constant terms which have been
omitted in the definition (\ref{eqn:defEI}) have to be restored in order to
establish a connection with the derivation algebra. In particular, the
translation of modified iterated Eisenstein integrals into the language of
non-commutative words built from letters $g_{k}$ described in
\subsecref{subsec:deralg} involves various shifts $\sim \zm_{k} g_0$, 
\begin{align}
&\psi\big[ \gamma_0(k_1, 0^{p_1},k_2, 0^{p_2},\ldots, k_n, 0^{p_n}) \big] = 
(-1)^{p_1+p_2+\ldots+p_n}   \\
&\ \ \ \   \times \  (g_0)^{p_n} \Big( \frac{ g_{k_n} }{k_n-1} - 2 \zm_{k_n} g_0 \Big) \cdots (g_0)^{p_2} \Big( \frac{ g_{k_2} }{k_2-1} - 2 \zm_{k_2} g_0 \Big)(g_0)^{p_1} \Big( \frac{ g_{k_1} }{k_1-1} - 2 \zm_{k_1} g_0 \Big) \, ,
\notag
\end{align}
where $(g_0)^n$ refers to $n$ adjacent letters $g_0$. Furthermore, the
concatenation of words is understood to act linearly,
e.g.~$g_2(\zm_4g_0+g_4)g_8=\zm_4g_2g_0g_8+g_2g_4g_8$.  Nevertheless, the
counting of indecomposable eMZVs remains unmodified when projecting to the
maximal component of their $\gm$-representation, see the discussion below
\eqn{eqn:translation1G}.


\section{Conclusions}

In this work we have been studying the systematics of relations between eMZVs.
Our results support the conjecture that the entirety of relations can be traced
back to reflection, shuffle and Fay identities. 

The numbers $N(\ell_\om,w_\om)$ of indecomposable eMZVs at any weight and
length can be explained once their connection to a special derivation algebra
is taken into account: Any eMZV can be expressed in terms of iterated integrals
over Eisenstein series whose appearance in turn is governed by the derivation
algebra. 

Our results for the numbers $N(\ell_\om,w_\om)$ of indecomposable eMZVs for
various weights $w_\om$ and lengths $\ell_\om$ are listed in table
\ref{tab:emzvbasisdim}. In addition, there are all-weight formul\ae{} available
for $\ell_\om\leq 4$ and odd values of $w_\om+\ell_\om$,
\begin{equation}
N(2,w_\om) = 1
 \co N(3,w_\om) = \left \lceil \frac{1}{6}w_\om \right \rceil
 \co N(4,w_\om) = \left \lfloor \frac{1}{2}+ \frac{1}{48}(w_\om+5)^2 \right \rfloor \ , 
\label{results}
\end{equation}
where the expression for $N(4,w_\om)$ is conjectural.  Because of the
diversity of constraints originating from the derivation algebra as described
in \secref{sec:gamma}, a closed formula for all weights and lengths is
challenging to find and not yet available. A closely related issue is the
computation of the dimensions of the Lie algebra $\DAlg$, which has been
carried out by Brown for depths $1$, $2$ and $3$~\cite{Brown:LetterToMatthes}.

Explicit $q$-expansions for eMZVs are accessible using a slightly modified
version of iterated Eisenstein integrals described in \subsecref{sec:simple}.
The resulting closed expression can be found in \eqn{qgamma2}.

The improved understanding of eMZVs raises a variety of follow-up questions,
starting with a connection of the underlying elliptic iterated integrals with
recent results on Feynman integrals \cite{Bloch:2013tra, Adams:2014vja,
Bloch:2014qca, Adams:2015gva}. In particular, the techniques which led to the
$q$-expansions of eMZVs furnish a convenient starting point to connect with the
functions $\ELi$ introduced in ref.~\cite{Adams:2014vja} and generalized in
ref.~\cite{Adams:2015gva}.

The appearance of eMZVs in one-loop scattering amplitudes of the open
superstring \cite{Broedel:2014vla} suggested a systematic study of
indecomposable eMZVs. The results of the current article should pave the way
towards a compact form of string corrections at higher orders in $\ap$ and
might even lead to a glimpse of an all-order pattern. The existence of such a
description is not unlikely: for open-string tree-level amplitudes a recursive
formula based on the Drinfeld associator is known. It was found by extending an
initial observation in ref.~\cite{Drummond:2013vz} into a recursive computation
of the complete $\ap$-expansions in ref.~\cite{Broedel:2013aza}. Similarly, the
$\ap$-expansion at one-loop might be accessible by using the elliptic
associators discussed in ref.~\cite{EnriquezEllAss}.

The $\ap$-expansion of the closed-string four-point amplitude at genus one has
been investigated in refs.~\cite{Green:1999pv, Green:2008uj, D'Hoker:2015foa},
see \cite{Richards:2008jg, Green:2013bza} for generalizations to five
external states. The functions appearing in those amplitudes
include non-holomorphic Eisenstein series and a variety of their
generalizations which have been analyzed in ref.~\cite{D'Hoker:2015foa}. It
would be interesting to establish a connection between these non-holomorphic
functions and modular-invariant combinations of eMZVs and their
counterpart originating from the other homology cycle.


\subsection*{Acknowledgments}

We are grateful to Carlos Mafra for collaboration in early stages of the
project and on related topics. We would like to thank Aaron Pollack for helpful
email correspondence regarding derivation algebras. Furthermore we are grateful
to Axel Kleinschmidt, Michael Green, Eric D'Hoker for helpful discussions.  In
particular, we would like to acknowledge two elaborate discussions with Francis
Brown in which he kindly explained to us -- among other things -- the
connection to ref.~\cite{Brown:mmv}. Moreover, we are indebted to an anonymous
referee for a variety of very valuable and profound suggestions on an earlier
version of this paper.

Furthermore, we would like to thank Michael Green for an invitation to
Cambridge and the Department of Applied Mathematics and Theoretical Physics of
the University of Cambridge for hospitality. We acknowledge financial support
by the European Research Council Advanced Grant No.~247252 of Michael Green.
JB and NM would like to thank the Albert-Einstein-Institute for hospitality.

Finally, we would like to acknowledge the Mainz Institute for Theoretical
Physics (MITP) for its hospitality and its partial support during completion of
this work. Moreover, we are grateful to the participants of the MITP workshop
``Amplitudes, Motives and Beyond'', in particular Herbert Gangl and Anton
Mellit, for numerous discussions about elliptic multiple zeta values and
related topics.

\section*{Appendix}
\appendix

\section{eMZV relations}


\subsection{Decomposition of boring eMZVs}
\label{app:boringeMZVs}

By \eqn{eqn:ellone}, all the above examples of shuffle-reductions of boring
eMZVs can be identified as special cases of the following general identity
\begin{align}
\omm(B)  \Big|_{\te{boring}} &= \sum_{k=1}^\infty D_{2k} \sum_{B  =A_1 A_2\ldots A_{2k} \atop{\omm(A_i) \ \te{interesting}}} \omm(A_1)\omm(A_2) \ldots  \omm(A_{2k}) \ ,
\label{decon3}
\end{align}
whose rational coefficients $D_{2}=\frac{1}{2}, \ D_4=-\frac{1}{8},\ldots$ are given by
\begin{equation}
D_{2k} = (-1)^{k-1} \frac{(2k-3)!!}{k! \, 2^k}   \ .
\label{ckexpl}
\end{equation}
%
The arguments $B\equiv n_1,n_2,\ldots, n_r$ of the boring eMZVs on the
left-hand side are deconcatenated\footnote{For example, the $k=1$ part of
  \eqn{decon3} encompasses those deconcatenations $B=A_1 A_2$ into $A_1 =
  n_1,n_2,\ldots,n_j$ and $A_2=n_{j+1},\ldots,n_r$ where
  $\omm(n_1,n_2,\ldots,n_j)$ and $\omm(n_{j+1},\ldots,n_r)$ are interesting
eMZVs.} into smaller tuples $A_j= a^j_1,a^j_2,\ldots,a^j_{m_j}$ such that
all eMZVs $\omm(A_j)$ are interesting. Only even numbers of interesting
$\omm(A_j)$ are compatible with the boring nature of $\omm(B)$, and the
concatenation $A_j A_{j+1}$ in \eqn{decon3} is defined to yield
$a^j_1,\ldots,a^j_{m_j},a^{j+1}_1,\ldots,a^{j+1}_{m_{j+1}}$.

Note that the first appearance of $D_4=-\frac{1}{8}$ can be seen from the
second case $\omm(n_1,n_2,n_3,n_4)$ in \eqn{sh42}. The vanishing of eMZVs with
all entries odd (cf.~\eqn{eqn:lm}) follows from the absence of deconcatenations
into tuples $A_j$ with $\omm(A_j)$ interesting.

In order to prove\footnote{We are grateful to an anonymous referee for suggesting the proof.} \eqns{decon3}{ckexpl}, we recall that the antipode
\begin{equation}
{\cal S}(n_1,n_2,\ldots ,n_r) \equiv (-1)^r (n_r,\ldots,n_2,n_1)
\label{prf1}
\end{equation}
in the shuffle algebra of words $B=n_1,n_2,\ldots, n_r$ satisfies the following defining property \cite{Abe}
\begin{equation}
B + {\cal S}(B) + \sum_{B = A_1 A_2 \atop{A_1,A_2 \neq \emptyset }} A_1 \shuffle {\cal S}(A_2) = 0  \ , \ \ \ \ B \neq \emptyset \ .
\label{prf2}
\end{equation}
Since boring and interesting eMZVs can be neatly characterized through the antipode \eqn{prf1},
\begin{equation}
\omm({\cal S}(B)) = \left\{ \begin{array}{rl}
\omm(B) &: \ \omm(B) \ \te{boring} \\
-\omm(B) &: \ \omm(B) \ \te{interesting} 
\end{array} \right.  \ ,
\label{prf3}
\end{equation}
applying $\omm(\cdot)$ to (\ref{prf2}) with boring $\omm(B)$ yields
\begin{equation}
\omm(B)  \Big|_{\te{boring}} = \frac{1}{2} \bigg\{  \! \! \! \! \sum_{B  =A_1 A_2 \atop{\omm(A_i) \ \te{interesting}}}   \! \! \! \!\omm(A_1)\omm(A_2) -  \! \! \! \! \sum_{B  =B_1 B_2 \atop{\omm(B_i) \ \te{boring}}}   \! \! \! \! \omm(B_1)\omm(B_2)
\bigg\} \ .
\label{prf4}
\end{equation}
This formula can be recursively applied to the boring factors $\omm(B_i)$ on
the right-hand side until only interesting contributions remain, leading to the
structure of \eqn{decon3}. The coefficients $D_{2k}$ in \eqn{ckexpl} are
determined by the combinatorics of iterating \eqn{prf4}, e.g.
\begin{align}
\omm(B)  \Big|_{\te{boring}} &= \frac{1}{2}  \! \! \! \! \! \! \sum_{B  =A_1 A_2 \atop{\omm(A_i) \ \te{interesting}}}  \! \! \! \! \! \!\omm(A_1)\omm(A_2)
 - \frac{1}{8}  \! \! \! \! \sum_{B  =B_1 B_2 \atop{\omm(B_i) \ \te{boring}}}  \! \! \!  \bigg\{   \! \! \! \sum_{B_1 =A_1 A_2 \atop{\omm(A_i) \ \te{interesting}}}  \! \! \! \! \! \! \omm(A_1)\omm(A_2) -  \! \! \! \! \sum_{B_1 =B_3 B_4 \atop{\omm(B_i) \ \te{boring}}}   \! \! \! \! \omm(B_3)\omm(B_4) \bigg\} \notag\\
 & \ \ \ \ \ \  \ \ \ \ \ \  \ \ \ \ \ \  \ \ \ \ \ \ \  \ \ \ \ \ \ \times  \bigg\{   \! \! \! \sum_{B_2 =A_3 A_4 \atop{\omm(A_i) \ \te{interesting}}}  \! \! \! \! \! \! \omm(A_3)\omm(A_4) -  \! \! \! \! \sum_{B_2 =B_5 B_6 \atop{\omm(B_i) \ \te{boring}}}   \! \! \! \! \omm(B_5)\omm(B_6) \bigg\}
\label{prf5}
\end{align}
reproduces the first two terms with $k\leq 2$ in \eqn{decon3}. In the next
iteration step towards $k=3$, either the first boring pair $\omm(B_3)
\omm(B_4)$ or the second one $\omm(B_5) \omm(B_6)$ can be rearranged via
\eqn{prf4}. Since both of them contribute $\frac{1}{32} \sum_{B =A_1\ldots A_6}
\omm(A_1) \ldots \omm(A_6)$ with $\omm(A_i)$ interesting, the above coefficient
$D_6=\frac{1}{16}$ rests on the two subdivisions of the schematic form $\{ (A_1
A_2)(A_3 A_4) \} (A_5 A_6)$ or $ (A_1 A_2) \{(A_3 A_4) (A_5 A_6) \}$, referring
to the application of \eqn{prf4} to either $\omm(B_3) \omm(B_4)$ or $\omm(B_5)
\omm(B_6)$, respectively.

The contributions of $\sum_{B =A_1\ldots A_{2k}}$ from the iteration of
\eqn{prf4} can be organized in terms of full binary trees with $k-1$ internal
vertices and $k$ leaves. Internal vertices represent the expansion of pairs of
boring eMZVs via \eqn{prf4} and at each of the $k$ leaves, a pair of
interesting eMZVs is kept. Hence, the coefficient of $\sum_{B =A_1\ldots
A_{2k}} \omm(A_1) \ldots \omm(A_{2k})$ in \eqn{decon3} is given by
\begin{equation}
D_{2k} = \frac{1}{2}  \left( -\frac{1}{4} \right)^{k-1} \cdot \#(\te{full binary trees with $k$ leaves})\ ,
\label{prf6}
\end{equation}
where each additional application of \eqn{prf4} to a pair of boring $\omm(B_i)$
involves a prefactor of $-\frac{1}{4}$. Finally, since full binary trees with
$k$ leaves are counted by the Catalan number $C_{k-1}$ \cite{EnumComb} with 
\begin{equation}
C_n = \frac{(2n)!}{(n+1)! \, n!}  = 2^n  \frac{ (2n-1)!!}{(n+1)!} \,,
\label{catalan}
\end{equation}
we recover the coefficients $D_{2k} =  \frac{ (-1)^{k-1} }{ 2^{2k-1} } \, C_{k-1} $ in \eqn{ckexpl} from \eqn{prf6}.


\subsection{More general Fay identities}
\label{app:moreFay}

The relation \eqn{extra21} among elliptic iterated integrals yields various Fay
identities in the limit $z\rightarrow 1$ and generalizes as follows to multiple
appearances of the argument among the labels:
\begin{align}
& \GLarg{n_1 &n_2 &\ldots &n_{k} &n_{k+1} &\ldots &n_r}{z &z &\ldots &z &0 &\ldots  &0}{z} = (-1)^k \zms( \underbrace{0 \ldots 0}_{r-k}  \underbrace{1 \ldots 1}_{k} )\prod_{j=1}^r \delta_{n_j,1} \notag
\\
&-(-1)^{n_k}\int^z_0 \dd t \, f^{(n_k+n_{k+1})}(t) \GLarg{n_1 &\ldots &n_{k-1} &0 &n_{k+2} &\ldots &n_r}{t &\ldots &t &0 &0&\ldots &0}{t}  \label{tzero81} \\
&+ \sum_{j=0}^{n_{k+1}} {n_k-1+j \choose j} \int^z_0 \dd t \, f^{(n_{k+1}-j)}(t) \GLarg{n_1 &\ldots &n_{k-1} &n_k+j &n_{k+2} &\ldots &n_r}{t &\ldots &t &t &0 &\ldots &0}{t} \notag \\
&+ \sum_{j=0}^{n_k} {n_{k+1}-1+j \choose j} (-1)^{n_k+j} \int^z_0 \dd t \, f^{(n_k-j)}(t) \GLarg{n_1&\ldots &n_{k-1} &n_{k+1}+j &n_{k+2} &\ldots &n_r }{t &\ldots &t &0 &0 &\ldots &0}{t}
\,.\notag
\end{align}
The MZV in the first line stems from the limit $z\to0$ of
$\Gamma(\underbrace{\begin{smallmatrix}1&\cdots&1\\z&\cdots&z\end{smallmatrix}}_k\underbrace{\begin{smallmatrix}1&\cdots&1\\0&\cdots&0\end{smallmatrix}}_{r-k};z)$,
where every $f^{(1)}(z)$ can be replaced by $\frac{1}{z}$ in this regime. As
explained in ref.~\cite{Broedel:2014vla}, the elliptic iterated integrals then
reduce to particular instances of multiple polylogarithms, which can be shown
to yield MZVs in this case.  

The first
novel eMZV relations follow from the limit $z\rightarrow 1$ of \eqn{tzero81} at
$k=2$ and $r=4,5$:
\begin{align}
 \GLarg{n_1 &n_2  &n_{3} &n_{4} }{z &z  &0 &0}{z} &= 
 -\frac{1}{4} \zm_{4} \delta_{n_1,1}\delta_{n_2,1}\delta_{n_3,1}\delta_{n_4,1}- 
 (-1)^{n_2} \int^z_0 \dd t \, f^{(n_2+n_3)}(t) \GLarg{n_1 &0 &n_4}{t &0 &0}{t}  \notag \\
 &
+ \sum_{j=0}^{n_3} {n_2-1+j \choose j} \int^z_0 \dd t \, f^{(n_3-j)}(t) \GLarg{n_1 &n_2+j &n_4}{t &t &0}{t} \label{comp1}\\
&+ \sum_{j=0}^{n_2} {n_3-1+j\choose j}  (-1)^{n_2+j} \int^z_0 \dd t \, f^{(n_2-j)}(t) \GLarg{n_1 &n_3+j &n_4}{t &0&0}{t}
\notag
\\
 \GLarg{n_1 &n_2  &n_{3} &n_{4} &n_{5}}{z &z  &0 &0 &0}{z} &=
 (2 \zm_{5} - \zm_{2}\zm_{3}) \Big(\prod_{j=1}^5 \delta_{n_j,1}\Big)  - (-1)^{n_2} \int^z_0 \dd t \, f^{(n_2+n_3)}(t) \GLarg{n_1 &0 &n_4 &n_5}{t &0&0&0}{t}
 \notag\\
& + \sum_{j=0}^{n_3} {n_2-1+j\choose j} \int^z_0 \dd t \, f^{(n_3-j)}(t) \GLarg{ n_1 &n_2+j &n_4 &n_5}{t&t&0&0}{t} \label{comp2} \\
& + \sum_{j=0}^{n_2}  {n_3-1+j\choose j} (-1)^{n_2+j} \int^z_0 \dd t \, f^{(n_2-j)}(t) \GLarg{n_1 &n_3+j &n_4 &n_5}{t &0&0&0}{t} \ . \notag
\end{align}
In particular, note that the product $\zm_{2}\zm_{3}$ is absent in
\eqn{extra21} at $r=5$. Also, note that the divergent nature of $f^{(1)}$
causes extra complications in the limit $z\rightarrow 1$ of \eqn{comp1} if
$n_i=1$ for $i=1,2,3,4$ and \eqn{comp2} if $n_2=n_3=n_4=1$ and one of
$n_1=1$ or $n_5=1$.

\section{Iterated Eisenstein integrals versus eMZVs: examples}
\label{app:conv}

In this appendix, we supplement further examples for the conversion of eMZVs
into modified iterated Eisenstein integrals as defined in \eqn{eqn:defEI}.


\subsection{Conversion of $\omm(0,0,\ldots,0,n)$}
\label{manyzeros}

For eMZVs with only one non-zero entry, a closed formula can be given for their
conversion into iterated Eisenstein integrals. At length $\ell_\om=4$ and
$\ell_\om=5$, \eqns{eqn:gamma1}{eqn:gamma2} can be generalized to
\begin{align}
\omm(0,0,0,n) &= \delta_{n,1} \left( \frac{i \pi }{12} + \frac{ \zeta_3}{4\pi^2} \right) +\frac{n}{3!} \gmz(n+1)+ n(n+1)(n+2) \gmz(n+3,0,0) 
\label{stan4}
\\
\omm(0,0,0,0,n) &= - \frac{2\zm_n}{5!}  - \frac{n}{3!} (n+1)\gmz(n+2,0)- n(n+1)(n+2)(n+3) \gmz(n+4,0,0,0) \ ,
\label{stan5}
\end{align}
where $n$ is chosen to be odd in \eqn{stan4} and even in \eqn{stan5}. At
arbitrary length $\ell$, we have
\begin{align}
\omm(\underbrace{0,0,\ldots,0}_{\ell-1},n) &= \left\{ 
\begin{array}{cl}\displaystyle
 \omm_0(0^{\ell-1},n)+
\sum_{i=1,3,5,\atop{\ldots,\ell-1}} \frac{ \gmz(n+i,0^{i-1})}{(\ell-i)!} \prod_{j=0}^{i-1} (n+j) &: \ \ell \ \te{even}, \ n \ \te{odd}
\\
\displaystyle
-\frac{2\zm_n}{\ell!} 
-\sum_{i=2,4,6,\atop{\ldots,\ell-1}} \frac{ \gmz(n+i,0^{i-1})}{(\ell-i)!} \prod_{j=0}^{i-1} (n+j) &: \ \ell \ \te{odd}, \ n \ \te{even} 
\end{array}
\right. \ ,
\label{anylength}
\end{align}
where the constant term for odd values of $n$ vanishes except for weight one,
\begin{equation}
\omm_0(0^{\ell-1},n) = \delta_{n,1}\bigg\{ \frac{ i\pi  }{2 (\ell-1)!} - \sum_{k=1}^{\lfloor\ell/2 \rfloor-1} \frac{(-1)^k}{\big(\ell - (2k+1)\big)!} \frac{ \zeta_{2k+1} }{(4\pi^2)^k} \bigg\}
\ . \label{constw1}
\end{equation}
The $q$-expansion of \eqn{anylength} can be inferred from the special case of \eqn{qgamma2},
\begin{equation}
\gmz(k,0^{p-1}) = - \frac{ 2 (2\pi i)^{k-2p} }{(k-1)!} \sum_{n=1}^{\infty} \frac{ \sigma_{k-1}(n) q^n }{n^p} \ ,
\end{equation}
see \eqn{qgamma3} for the definition of the divisor sum $\sigma_k(n)$.

\subsection{Conversion of indecomposable eMZVs at $\ell_\om \geq 3$}
\label{lesszeros}

Among the indecomposable eMZVs beyond $\omm(0,\ldots,0,n) $, the simplest case
$\omm(0,3,5)$ is converted to (modified) iterated Eisenstein integrals in
\eqns{eqn:w035result}{forapp}.  Beyond that, we find for example
\begin{align}
\omm(0,3,7) &= -1848 \gm(12,0) -294 \gm(8,4) + \te{nmt}\nnl
            &= -1848 \gmz(12,0) -294 \gmz(8,4) - 75 \big(\!\gmz(6)\big)^2 
	       + 588 \zm_4 \gmz(8,0) -504 \zm_8 \gmz(4,0)
\nnl
\omm(0,3,9) &=
-5616 \gm(14,0) -729 \gm(10,4) -315 \gm(8,6) + \te{nmt}\nnl
            &= -5616 \gmz(14,0) -729 \gmz(10,4) -315 \gmz(8,6) -210 \gmz(6)\gmz(8)\nnl
            &\quad+1458 \zm_4\gmz(10,0) + 630 \zm_6 \gmz(8,0)-630 \zm_6 \gmz(6,0) -1350 \zm_{10} \gmz(4,0) 
\notag \\
\omm(0,3,11) &=
-13695 \gm(16,0)-1452 \gm(12,4)-990 \gm(10,6)+ \te{nmt}\nnl
&=-13695 \gmz(16,0)-1452 \gmz(12,4)-\frac{735}{2} \big(\!\gmz(8)\big)^2-990 \gmz(10,6)-270 \gmz(6)\gmz(10)\nnl
&\quad +2904 \zm_4 \gmz(12,0) + 1980 \zm_6 \gmz(10,0) -1980 \zm_{10} \gmz(6,0) -2772 \zm_{12}\gmz(4,0) 
\nnl
\omm(0,5,9) &=
-30105 \gm(16,0)-5445 \gm(12,4)-3105 \gm(10,6)+\te{nmt}\nnl
&= -30105 \gmz(16,0)-5445 \gmz(12,4)-3105 \gmz(10,6)-\frac{735}{2} \big(\!\gmz(8)\big)^2\nnl
&\quad +10890 \zm_4 \gmz(12,0) + 6210 \zm_6 \gmz(10,0) - 5850 \zm_{10} \gmz(6,0) - 8910 \zm_{12}
\gmz(4,0) 
\end{align}
at length three, and 
\begin{align}
\omm(0,0,2,3) &=
  252 \gm(8,0,0)-18 \gm(4,4,0)+\frac{5}{6} \gm(6)+ \te{nmt}\nnl
&=252 \gmz(8,0,0)-18 \gmz(4,4,0)+\frac{5}{6} \gmz(6)-72 \zm_4 \gmz(4,0,0)
\notag \\
\omm(0,0,2,5) &=
  2826 \gm(10,0,0)+150 \gm(6,4,0)+180 \gm(6,0,4)+\frac{7}{6} \gm(8)+ \te{nmt}\nnl
&=2826 \gmz(10,0,0)+150 \gmz(6,4,0)+180 \gmz(6,0,4)+\frac{7}{6} \gmz(8)\nnl
&\quad - 660 \zm_4 \gmz(6,0,0) + 180 \zm_6 \gmz(4,0,0)
\notag \\
\omm(0,0,4,3) &=
-2340 \gm(10,0,0) -300 \gm(6,4,0) -120 \gm(6,0,4)  +\frac{7}{6} \gm(8)+ \te{nmt}\nnl
&= -2340 \gmz(10,0,0) -300 \gmz(6,4,0) -120 \gmz(6,0,4) -60 \gmz(4)\gmz(6,0) +\frac{7}{6} \gmz(8)\nnl
&\quad +480 \zm_4 \gmz(6,0,0) -1080 \zm_6 \gmz(4,0,0) -3 \zm_4 \gmz(4)
\end{align}
at length four, where ``nmt'' refers to non-maximal terms as explained after
\eqn{eqn:translation1G}.
The $q$-expansion of the constituents is given by \eqn{qgamma2}.


\section[Examples for relations in the derivation algebra $\DAlg$]
{Examples for relations in the derivation algebra \texorpdfstring{$\DAlg$}{u}}
\subsection{Known relations}
\label{sec:knownrelations}

Irreducible relations $r_{\text{weight}}^{\text{depth}}$ are listed in table
\ref{tab:irreducible}. For depth two, all relations can be obtained from
\eqn{closedpoll}. At depth three, we can confirm the relations listed in
\eqn{poll4} as well as~\cite{Pollack}:
\begin{align}
  r^3_{20}:\quad 0&= 1050 [\ep_0,[\ep_6,\ep_{14}]]-6580 [\ep_0,[\ep_8,\ep_{12}]]+4320 [\ep_4,[\ep_0,\ep_{16}]]-10970 [\ep_4,[\ep_4,\ep_{12}]]\nnl
  & +166675 [\ep_4,[\ep_6,\ep_{10}]]-17150 [\ep_6,[\ep_0,\ep_{14}]]-500675 [\ep_6,[\ep_6,\ep_8]]+30184 [\ep_8,[\ep_0,\ep_{12}]]\nnl
  & +80388 [\ep_8,[\ep_4,\ep_8]]-17325 [\ep_{10},[\ep_0,\ep_{10}]]
\end{align}
\begin{align}
r^3_{22}:\quad 0&= 40 [\ep_0,[\ep_6,\ep_{16}]]-280 [\ep_0,[\ep_8,\ep_{14}]]+910 [\ep_0,[\ep_{10},\ep_{12}]]-360 [\ep_4,[\ep_0,\ep_{18}]]\nnl
& -11535 [\ep_4,[\ep_6,\ep_{12}]]+6069 [\ep_4,[\ep_8,\ep_{10}]]+1320 [\ep_6,[\ep_0,\ep_{16}]]+15140 [\ep_6,[\ep_4,\ep_{12}]]\nnl
& -7150 [\ep_6,[\ep_6,\ep_{10}]]-1820 [\ep_8,[\ep_0,\ep_{14}]]-12922 [\ep_8,[\ep_6,\ep_8]]+858 [\ep_{10},[\ep_0,\ep_{12}]]
\end{align}
\begin{align}
r^4_{18}:\quad 0&= [\ep_0,[\ep_0,[\ep_6,\ep_{12}]]]-\frac{215}{74} [\ep_0,[\ep_0,[\ep_8,\ep_{10}]]]-\frac{2323}{518} [\ep_0,[\ep_4,[\ep_6,\ep_8]]]+\frac{218}{37} [\ep_0,[\ep_6,[\ep_4,\ep_8]]]\nnl
& +\frac{60}{407} [\ep_4,[\ep_0,[\ep_0,\ep_{14}]]]+\frac{285561}{5698} [\ep_4,[\ep_0,[\ep_6,\ep_8]]]+\frac{8599}{1628} [\ep_4,[\ep_4,[\ep_0,\ep_{10}]]]\nnl
& +\frac{53855}{444} [\ep_4,[\ep_4,[\ep_4,\ep_6]]]-\frac{691}{333} [\ep_6,[\ep_0,[\ep_0,\ep_{12}]]]-\frac{19853}{518} [\ep_6,[\ep_0,[\ep_4,\ep_8]]]\nnl
& -\frac{691}{74} [\ep_{10},[\ep_0,[\ep_0,\ep_8]]]+\frac{691}{111} [\ep_{12},[\ep_0,[\ep_0,\ep_6]]]-\frac{60}{37} [\ep_{14},[\ep_0,[\ep_0,\ep_4]]]-\frac{87595}{1554} [\ep_6,[\ep_4,[\ep_0,\ep_8]]]
\nnl
& +\frac{17275}{333} [\ep_6,[\ep_6,[\ep_0,\ep_6]]]+\frac{3455}{518} [\ep_8,[\ep_0,[\ep_0,\ep_{10}]]]+\frac{49565}{518} [\ep_8,[\ep_0,[\ep_4,\ep_6]]]
\end{align}
\begin{align}
r^4_{22}:\quad 0 &= [\ep_0,[\ep_0,[\ep_8,\ep_{14}]]]+\frac{192903}{230} [\ep_0,[\ep_4,[\ep_6,\ep_{12}]]]-\frac{861492}{805} [\ep_0,[\ep_6,[\ep_4,\ep_{12}]]]\nnl
& +\frac{134488}{161} [\ep_0,[\ep_6,[\ep_6,\ep_{10}]]]+\frac{6588}{805} [\ep_4,[\ep_0,[\ep_0,\ep_{18}]]]+\frac{269217}{805} [\ep_4,[\ep_0,[\ep_6,\ep_{12}]]]\nnl
& -\frac{39418}{115} [\ep_4,[\ep_0,[\ep_8,\ep_{10}]]]-\frac{13253}{115} [\ep_4,[\ep_4,[\ep_0,\ep_{14}]]]-\frac{18221}{115} [\ep_4,[\ep_4,[\ep_6,\ep_8]]]\nnl
& +\frac{33109}{322} [\ep_6,[\ep_0,[\ep_6,\ep_{10}]]]+\frac{25095129}{37375} [\ep_6,[\ep_4,[\ep_0,\ep_{12}]]]+\frac{11266827}{5750} [\ep_6,[\ep_4,[\ep_4,\ep_8]]]\nnl
& -\frac{786557}{644} [\ep_6,[\ep_6,[\ep_0,\ep_{10}]]]+\frac{80233}{1265} [\ep_8,[\ep_0,[\ep_0,\ep_{14}]]]+\frac{21742068}{6325} [\ep_8,[\ep_0,[\ep_6,\ep_8]]]\nnl
& -\frac{112835}{253} [\ep_8,[\ep_4,[\ep_0,\ep_{10}]]]+\frac{403764}{115} [\ep_8,[\ep_4,[\ep_4,\ep_6]]]+\frac{644938}{575} [\ep_8,[\ep_6,[\ep_0,\ep_8]]]\nnl
& -\frac{103859}{115} [\ep_{10},[\ep_0,[\ep_4,\ep_8]]]+\frac{301851}{8050} [\ep_{12},[\ep_0,[\ep_0,\ep_{10}]]]+\frac{734133}{805} [\ep_{12},[\ep_0,[\ep_4,\ep_6]]]\nnl
& -\frac{493889}{8050} [\ep_{10},[\ep_0,[\ep_0,\ep_{12}]]]-\frac{372888}{10465} [\ep_6,[\ep_0,[\ep_0,\ep_{16}]]]-\frac{23054063}{52325} [\ep_6,[\ep_0,[\ep_4,\ep_{12}]]]  \nnl
& -\frac{1015637}{1150} [\ep_0,[\ep_4,[\ep_8,\ep_{10}]]]-\frac{27458211}{3220} [\ep_6,[\ep_6,[\ep_4,\ep_6]]]-\frac{23679}{8050} [\ep_0,[\ep_0,[\ep_{10},\ep_{12}]]]\nnl
& -\frac{1913}{115} [\ep_{14},[\ep_0,[\ep_0,\ep_8]]]+\frac{672}{115} [\ep_{16},[\ep_0,[\ep_0,\ep_6]]]-\frac{972}{805} [\ep_{18},[\ep_0,[\ep_0,\ep_4]]] 
\end{align}

\subsection{New relations}
\label{sec:explrel}
At depth 5 we explicitly isolated the irreducible relation $r^5_{20}$, which is apparently new:  
\begin{align}
  r^5_{20}:\quad 0&=2206388620800\,[\ep_0,[\ep_0,[\ep_0,[\ep_4,\ep_{16}]]]]-8366188740000\,[\ep_0,[\ep_0,[\ep_0,[\ep_6,\ep_{14}]]]]\nnl
&+12305858292000\,[\ep_0,[\ep_0,[\ep_0,[\ep_8,\ep_{12}]]]]-1834700544000\,[\ep_0,[\ep_4,[\ep_0,[\ep_0,\ep_{16}]]]]\nnl
&+35687825530800\,[\ep_0,[\ep_4,[\ep_0,[\ep_4,\ep_{12}]]]]-109425220173750\,[\ep_0,[\ep_4,[\ep_0,[\ep_6,\ep_{10}]]]]\nnl
&-39970750599360\,[\ep_0,[\ep_4,[\ep_4,[\ep_0,\ep_{12}]]]]-380488416808500\,[\ep_0,[\ep_4,[\ep_4,[\ep_4,\ep_8]]]]\nnl
&+13171256280000\,[\ep_0,[\ep_6,[\ep_0,[\ep_0,\ep_{14}]]]]+220479512028750\,[\ep_0,[\ep_6,[\ep_0,[\ep_4,\ep_{10}]]]]\nnl
&-498847136287500\,[\ep_0,[\ep_6,[\ep_0,[\ep_6,\ep_8]]]]+220479512028750\,[\ep_0,[\ep_6,[\ep_4,[\ep_0,\ep_{10}]]]]\nnl
&-458212979593200\,[\ep_0,[\ep_6,[\ep_4,[\ep_4,\ep_6]]]]+17540335312500\,[\ep_0,[\ep_6,[\ep_6,[\ep_0,\ep_8]]]]\nnl
&-34407225652800\,[\ep_0,[\ep_8,[\ep_0,[\ep_0,\ep_{12}]]]]-97419791414400\,[\ep_0,[\ep_8,[\ep_0,[\ep_4,\ep_8]]]]\nnl
&-197536749664800\,[\ep_0,[\ep_8,[\ep_4,[\ep_0,\ep_8]]]]+22970739577500\,[\ep_0,[\ep_{10},[\ep_0,[\ep_0,\ep_{10}]]]]\nnl
&+161385266688750\,[\ep_0,[\ep_{10},[\ep_0,[\ep_4,\ep_6]]]]+611566848000\,[\ep_4,[\ep_0,[\ep_0,[\ep_0,\ep_{16}]]]]\nnl
&-58836403790864\,[\ep_4,[\ep_0,[\ep_0,[\ep_4,\ep_{12}]]]]+134572047805000\,[\ep_4,[\ep_0,[\ep_0,[\ep_6,\ep_{10}]]]]\nnl
&+965866444426884\,[\ep_4,[\ep_0,[\ep_4,[\ep_4,\ep_8]]]]+92810063342256\,[\ep_4,[\ep_4,[\ep_0,[\ep_0,\ep_{12}]]]]\nnl
&-204658497503460\,[\ep_4,[\ep_4,[\ep_0,[\ep_4,\ep_8]]]]+541534390897500\,[\ep_4,[\ep_4,[\ep_4,[\ep_0,\ep_8]]]]\nnl
&-215755493216250\,[\ep_4,[\ep_6,[\ep_0,[\ep_0,\ep_{10}]]]]-1490371718737200\,[\ep_4,[\ep_6,[\ep_0,[\ep_4,\ep_6]]]]\nnl
&+1032598095322950\,[\ep_4,[\ep_6,[\ep_4,[\ep_0,\ep_6]]]]+298655975581600\,[\ep_4,[\ep_8,[\ep_0,[\ep_0,\ep_8]]]]\nnl
&-220479512028750\,[\ep_4,[\ep_{10},[\ep_0,[\ep_0,\ep_6]]]]+54837332264496\,[\ep_4,[\ep_{12},[\ep_0,[\ep_0,\ep_4]]]]\nnl
&-6941740260000\,[\ep_6,[\ep_0,[\ep_0,[\ep_0,\ep_{14}]]]]-220479512028750\,[\ep_6,[\ep_0,[\ep_0,[\ep_4,\ep_{10}]]]]\nnl
&+231883232831250\,[\ep_6,[\ep_0,[\ep_0,[\ep_6,\ep_8]]]]+519528504682200\,[\ep_6,[\ep_0,[\ep_4,[\ep_4,\ep_6]]]]\nnl
&-220479512028750\,[\ep_6,[\ep_4,[\ep_0,[\ep_0,\ep_{10}]]]]-2120947122294000\,[\ep_6,[\ep_4,[\ep_0,[\ep_4,\ep_6]]]]\nnl
&-1538522546497950\,[\ep_6,[\ep_4,[\ep_4,[\ep_0,\ep_6]]]]+249423568143750\,[\ep_6,[\ep_6,[\ep_0,[\ep_0,\ep_8]]]]\nnl
&-266963903456250\,[\ep_6,[\ep_8,[\ep_0,[\ep_0,\ep_6]]]]+23162632092600\,[\ep_8,[\ep_0,[\ep_0,[\ep_0,\ep_{12}]]]]\nnl
&+184988881773150\,[\ep_8,[\ep_0,[\ep_0,[\ep_4,\ep_8]]]]+310347440367510\,[\ep_8,[\ep_4,[\ep_0,[\ep_0,\ep_8]]]]\nnl
&-183822714075000\,[\ep_8,[\ep_6,[\ep_0,[\ep_0,\ep_6]]]]+171943360038450\,[\ep_8,[\ep_8,[\ep_0,[\ep_0,\ep_4]]]]\nnl
&-22551859687500\,[\ep_{10},[\ep_0,[\ep_0,[\ep_0,\ep_{10}]]]]-240755752121625\,[\ep_{10},[\ep_0,[\ep_0,[\ep_4,\ep_6]]]]\nnl
&-104628710038125\,[\ep_{10},[\ep_4,[\ep_0,[\ep_0,\ep_6]]]]-14987648446875\,[\ep_{10},[\ep_6,[\ep_0,[\ep_0,\ep_4]]]]\nnl
&+11918038532400\,[\ep_{12},[\ep_0,[\ep_0,[\ep_0,\ep_8]]]]+46293152724000\,[\ep_{12},[\ep_4,[\ep_0,[\ep_0,\ep_4]]]]\nnl
&-8366188740000\,[\ep_{14},[\ep_0,[\ep_0,[\ep_0,\ep_6]]]]\,.
\label{eqn:newlength5}
\end{align}
The complete set of all irreducible relations known to us is available from
\begin{equation*}
  \text{\EMZVDatamine{}} \ ,
\end{equation*}
and all of them have been verified by evaluating the action on the letters
$x,y$ via \eqn{eqn:basicrel}.

\bibliographystyle{nb}
\bibliography{\jobname}

\end{document}